\documentclass[11pt,a4paper]{article}
\usepackage{jheppub}
\usepackage{graphicx}
\usepackage{dcolumn}
\usepackage{bm}
\usepackage{hyperref}
\usepackage{ragged2e} 
\usepackage{booktabs}
\usepackage[english]{babel}
\usepackage{amsmath,amssymb,amsbsy,amstext, amsthm, simplewick}
\usepackage{graphicx}
\usepackage{amsfonts}
\usepackage{amssymb}
\usepackage{upgreek}
\usepackage{cancel}
\usepackage{color}
\usepackage{slashed}
\usepackage[dvipsnames]{xcolor}
\usepackage{tikz-cd}
\usepackage{feynmp-auto}
\usepackage{color}
\usepackage{soul}
\usepackage{dsfont}
\usepackage{verbatim}
\usepackage[utf8]{inputenc} 
\usepackage{soul}
\usepackage{epigraph}
\usepackage{todonotes}
\usepackage{csquotes}
\usepackage{tikz}
\usepackage{subfig}

\newcommand{\half}{\frac{1}{2}}

\renewcommand{\c}[1]{\mathcal{#1}}

\newcommand{\sutt}{\left(SU(3)_C \times SU(3)_H\right)/\mathbb{Z}_3}
\newcommand{\suto}{\left(SU(3)_C \times U(1)_H\right)/\mathbb{Z}_3}

\def\@fnsymbol#1{\ensuremath{\ifcase#1\or $\Re$\or $\Im$\or  \else\@ctrerr\fi}}

\def\beq{\begin{equation}}
\def\eeq{\end{equation}}
\def\bea{\begin{eqnarray}}
\def\eea{\end{eqnarray}}

\newcommand\supsetsim{\mathrel{\substack{
  \textstyle\supset\\[-0.2ex]\textstyle\sim}}}

\begin{document}

  \title{Non-Invertible Peccei-Quinn Symmetry and the \\ Massless Quark Solution to the Strong CP Problem}

\author[a,b]{Clay C\'{o}rdova,}
\author[c]{Sungwoo Hong,}
\author[a,d]{Seth Koren}
\affiliation[a]{Enrico Fermi Institute, University of Chicago}
\affiliation[b]{Kadanoff Center for Theoretical Physics, University of Chicago}
\affiliation[c]{Department of Physics, Korea Advanced Institute of Science and Technology}
\affiliation[d]{Department of Physics and Astronomy, University of Notre Dame}

\abstract{
We consider theories of gauged quark flavor and identify non-invertible Peccei-Quinn symmetries arising from fractional instantons when the resulting gauge group has non-trivial global structure.
Such symmetries exist solely because the Standard Model has the same numbers of generations as colors, $N_g = N_c$. 
This leads us to a massless down-type quark solution to the strong CP problem in an ultraviolet $SU(9)$ theory of quark color-flavor unification. 
We show how the CKM flavor structure and weak CP violation can be generated without upsetting our solution. 
}
\maketitle

\section{Introduction} \label{sec:background}

We are now a hundred years removed from the birth of quantum physics, and who knows how long from its maturity. Yet we have already enjoyed a century of knowing the true mechanical laws which govern our universe. What can we say now of the program to understand the fundamental physical nature of the matter from which we are composed? 
The top-line conclusion is that the Standard Model of particle physics is a beautifully simple theory and represents a truly remarkable success of reductionism. It is astounding that we can understand the structure of the universe down to distances $\sim 10^{-19} \ \text{m}$, and the Standard Model description works exceedingly well.

The subleading story is that the structure of the Standard Model itself remains mysterious, and physicists have long been interested in pushing the program of reductionism even further: We wish to uncover a theory which explains why we exist at this point in the Standard Model's parameter space. There are some couple dozen numbers which must be inputted, including angles taking many different values, and some ratios of scales which are very large. Past general qualitative considerations (c.f.\ Dirac \cite{Dirac:1938mt}), the structure of quantum field theory characterizes how generally difficult it is to have an ultraviolet theory which explains the variety of needed numbers (c.f.\ 't Hooft \cite{tHooft:1979rat}), and the SM poses certain challenges. 

The multiplicity of inputs arises mainly from the yukawa sector, where three separate generations of matter means there are many masses and mixing angles. The yukawa couplings themselves have the benefit that they are `technically natural', which suggests that the puzzle of their sizes may be solved at small distances. The couplings $y_u, y_d, y_e$ (and $y_\nu$ with neutrino masses) are the sole breakings of the $U(3)^5$ flavor symmetries in three independent directions, and so a spurion analysis tells us in the SM they evolve with scale proportionally to themselves $\delta y_i \propto y_i$. This means we may hope to begin with (some of) these as exact symmetries and produce the needed sizes of yukawa couplings at some high scale, without this solution being disrupted by low-energy physics. 

Indeed this is what we have recently done in \cite{Cordova:2022fhg}, focusing on the technically-natural-but-ludicrously-small Dirac neutrino yukawas in the low-energy SM with right-handed neutrinos, where $y_{\nu_3}/y_{\tau} \lesssim 10^{-11}$ comparing the heaviest charged and neutral leptons.\footnote{Our analysis also revealed a way to generate Majorana neutrino masses by nonperturbative quantum effects.} 
With only the SM gauge group, the neutrino yukawas are spurions for a familiar, invertible symmetry. 
But an analysis of generalized symmetries in the lepton sector showed us that the neutrino yukawas can be protected by a non-invertible symmetry in lepton flavor gauge theories like $U(1)_{L_\mu - L_\tau}$.
This subtle interplay between the physics of lepton flavor monopoles and neutrino masses then guided us to a theory which produces these small numbers and is fully Dirac natural. 

That is, we wrote down an ultraviolet $SU(3)_H$ theory wherein all global symmetries are either good classical symmetries or explicitly broken by $\mathcal{O}(1)$ numbers like $y_\tau$.  
Then instanton effects of the ultraviolet theory break a classical symmetry and so produce naturally small Dirac neutrino yukawas from a Dirac natural theory $y_\nu \sim y_\tau \exp(-8\pi^2/g_H^2)$.
In understanding this model we effectively began a roadmap of model-building using non-invertible symmetries by showing how a technically natural, invertible spurion could be upgraded to a non-invertible spurion and then given a fully natural origin.

But the Standard Model also contains parameters which do not enjoy the protection of technical naturalness, and there is one such problem at each mass dimension $0, 2, 4$. The symmetry which could have protected the cosmological constant (CC), supersymmetry, is broken at a scale at least $\Lambda_{\scriptscriptstyle \rm SUSY}/\Lambda_{\scriptscriptstyle \rm CC} \gtrsim \text{TeV}/\text{meV} \sim 10^{15}$ larger than the energy scale of the CC, and presents a deep challenge to understanding the far infrared in our normal framework of Wilsonian effective field theory. Likewise the Higgs mass $\mu^2 H^\dagger H$ has no protective symmetry in the SM, which indicates the danger of the electroweak hierarchy problem---first realized as the doublet-triplet splitting problem of grand unification, but an important issue more generally in any UV field theory which introduces additional scales giving rise to $\delta \mu^2 \propto \Lambda^2$.  Finally the CP-violating phase in the strong sector, $\bar \theta$, is renormalized by the other source of quark CP-violation, the CKM angle $\delta_{\rm \scriptscriptstyle CKM}$. This lack of technical naturalness signals the strong CP problem.

One may na\"{i}vely wonder if generalized symmetries can be useful for these more serious naturalness problems of the SM where we cannot rely on the low energy theory being technically natural.
In this paper we exhibit this is indeed the case: As we will discuss below, the protective symmetries of the strong CP problem are quite subtle in the SM, and we will find that a non-invertible symmetry can shed light on a UV model which is fully Dirac natural and the smallness of the strong CP angle is explained. 
Admittedly, strong CP is special and the strong phase is closely related to the technically natural parameters of the yukawa sector, so we do not wish to give the impression that solving the big, dimensionful hierarchy problems is close at hand. Yet it is intriguing that we can go further than one might expect, and we do not know where else generalized symmetries may lead us. For other applications of generalized global symmetries in particle physics see e.g.\  \cite{Cordova:2018cvg,Wan:2019gqr,Davighi:2019rcd,Wang:2020xyo, Brennan:2020ehu,Hidaka:2020izy,Fan:2021ntg,Anber:2021upc,Wang:2021ayd,Wang:2021vki,Wang:2021hob,McNamara:2022lrw,Cordova:2022fhg,Cordova:2022qtz,Choi:2022rfe,Wang:2022eag,Yokokura:2022alv,Brennan:2023kpw,Cordova:2023her,Choi:2023pdp,vanBeest:2023dbu,Brennan:2023tae,Choi:2022fgx,Cordova:2023ent,Reece:2023iqn, Putrov:2023jqi, vanBeest:2023mbs,Aloni:2024jpb}, while for recent reviews we refer to \cite{Brennan:2023mmt, Bhardwaj:2023kri, Gomes:2023ahz, Luo:2023ive, Shao:2023gho}.

In this work we examine the generalized symmetries of the quark sector, and identify a non-invertible symmetry in the Standard Model when extended by a gauged horizontal symmetry of quarks which has non-trivial global structure.
This will require a more sophisticated generalized symmetry than our earlier work, where the non-invertible symmetry structure involved a $U(1)^{(1)}$ magnetic one-form symmetry, and  could be located from familiar triangle anomaly computations.
Here the non-invertible symmetry will connect a discrete $\mathbb{Z}_3^{(1)}$ magnetic one-form symmetry with a zero-form symmetry of quarks, and uncovering this structure will require us to examine our field theory on $S^2 \times S^2$ to see the relevant fractional instantons.
The payoff will be to discover a generalized symmetry which exists specifically because the SM bears out $N_c = N_g$, with the same number of colors as generations. This will point us to a special color-flavor unified theory in which we can construct a `massless down-type quark' solution to the strong CP problem using small instantons.

It is remarkable that such theories of non-trivial gauge-flavor unification are possible with the SM structure \cite{Allanach:2021bfe}, and intense study of their full phenomenologies is surely merited. We attempt to factorize the various physics effects here---for reasons we will discuss later on, slightly richer structure is needed to land precisely on SM flavor, and we postpone a fuller discussion to future work. Then our main goal is to understand and explain the intriguing non-perturbative physics such models possess, as informed by the generalized symmetry analysis, and how such models can solve the strong CP problem. We hope our work motivates increased activity on the perturbative structure of these theories, including detailed flavor model-building, and also the consideration of complementary signatures such as their early universe cosmologies.

First, in the rest of this introduction we will briefly review the strong CP problem as well as the original massless up quark solution to remind and orient readers on how to think about this style of solution. 
Then in Section \ref{sec:fromIR} we lay out the generalized symmetry analysis in the infrared theories of gauged quark flavor. 
In Section \ref{sec:fromUV} we introduce the ultraviolet model of $SU(9)$ color-flavor unification, how it solves the strong CP problem in the UV, and how this can be preserved while generating the CKM matrix. 
Various further directions of investigation are discussed in Section \ref{sec:generalize}.

Some technical and pedagogical material is reserved for the appendices. In Appendix \ref{app:fractionalInstanton} we review the construction of fractional instantons. In Appendix \ref{app:globalStructure} we review the global structure of gauge groups. In Appendix \ref{app:SU9thooft} we discuss one-loop corrections to 't Hooft vertices. 

\subsection{The Strong CP Problem}

The strong CP angle is the field-redefinition-invariant CP-odd phase combining the color topological density with the quark yukawa determinants,
\begin{equation} \label{eqn:bartheta}
    \bar \theta = \arg e^{-i\theta} \det\left(y_u y_d\right), 
\end{equation}
where $i \theta/32\pi^2$ is the coefficient of $G\tilde{G}$ in the Lagrangian, and if $\bar \theta = 0$ this means there is a field basis in which the quark yukawas have fully real eigenvalues. 
Impressive searches for, and constraints on, the electric dipole moment of the neutron indeed give us the upper bound  $\bar \theta \lesssim 10^{-10}$ \cite{Dar:2000tn,Abel:2020pzs}. 
If $\bar \theta$ were the only CP-violating parameter in the theory of quarks, then it would be technically natural, as spurious CP symmetry would imply small $\bar \theta$ is stable under RG flow $\delta \bar \theta \propto \bar \theta$. 

Alas, there is more CP-violation allowed in the quark yukawas for $N_g \geq 3$ generations.
The other CP-violating parameter is the `CKM phase' $\delta_{\rm \scriptscriptstyle CKM}$ which arises from the misalignment of the quark yukawas with the weak interaction basis, and which can be invariantly parametrized by the other CP-odd field-redefinition-invariant combination of the yukawas, the Jarlskog invariant\footnote{Throughout this work we will abuse terminology in referring to $\tilde{J}$ as `the Jarkskog invariant'. In fact this differs from the original definition by Jarlskog \cite{Jarlskog:1985ht} by a real proportionality factor. Explicitly, Jarlskog defined $i C \equiv \left[ y_u^\dagger y_u, y_d^\dagger y_d\right]$ and showed that 
\begin{equation*}
\begin{array}{llll}
     &  \det C = - 2 F F^\prime J, \\
     &  F = (m_t - m_c) (m_t - m_u) (m_c - m_u) / m_t^3, \\
     & F^\prime = (m_b - m_s) (m_b - m_d) (m_s - m_d) / m_b^3, \\
     & J = \text{Im} \left( V_{ij} V_{kl} V_{kj}^* V_{il}^* \right) \propto \sin \delta_{\rm \scriptscriptstyle CKM}
\end{array}  
\end{equation*}
where $V$ is the CKM matrix and $\delta_{\rm \scriptscriptstyle CKM}$ is the CP-odd phase. While $J$ sensibly splits off the observed dependence on the mass eigenvalues, and has been measured as $J \approx 3 \times 10^{-5}$ \cite{ParticleDataGroup:2022pth}, $\tilde{J} \equiv \det C$ can be written simply in terms of the yukawas and so is an easier diagnostic of nonvanishing CP-violation.}, 
\begin{equation}
    \tilde{J} = \text{Im} \det \left[y_u^\dagger y_u, y_d^\dagger y_d\right].
\end{equation}
The size of $\tilde{J}$ is often understood by going to the mass basis where the yukawas are diagonal and quark mixing lives in the CKM matrix, and parametrizing the CKM matrix in terms of mixing angles and a single CP-violating phase, $\delta_{\rm \scriptscriptstyle CKM} \sim 1.14$ \cite{ParticleDataGroup:2022pth}. This parameter is responsible for the CP-violation observed originally in decays of neutral kaons \cite{Christenson:1964fg}.

Then while the CKM phase obeys Dirac's naturalness principle in being an $\mathcal{O}(1)$ angle, $\bar \theta$ is not even technically natural, and in the limit of small $\bar \theta$ its renormalization group evolution goes as 
\begin{equation} \label{eqn:thetarunningSM}
    \delta \bar \theta \propto c \delta_{\rm\scriptscriptstyle CKM},
\end{equation}
so it seems there is no protection of $\bar \theta$ against quantum mechanical destabilization.
Now it is numerically true that the coefficient $c$, in the Standard Model, is very small \cite{Ellis:1978hq} as a result of the many approximate symmetries which we will discuss further in the next section. So given only the SM content, the running from \eqref{eqn:thetarunningSM} does not result in an infrared $\bar \theta$ in excess of the constraints. 

Still, the qualitative difference of $\bar \theta$ already being not the only CP-odd spurion indicates to us as effective field theorists that this can be a general concern in further UV theories. Indeed it is very easy for ultraviolet field theories to allow operators which would quickly introduce additional CP-violating phases into the quark sector, and would destabilize $\bar \theta$ far worse than does $\delta_{\rm\scriptscriptstyle CKM}$. And since $\bar \theta$ already is not the only CP-violating parameter, the obvious opportunity of simply imposing a CP symmetry on the ultraviolet of the theory is not straightforwardly available to us.

This naturalness tension has motivated decades of work on how physics beyond the Standard Model might square these two facts. 
One approach is to forge ahead by imposing a discrete symmetry in the ultraviolet (e.g. CP \`{a} la Nelson-Barr \cite{Nelson:1983zb,Barr:1984qx,Bento:1991ez}, or parity \cite{Barr:1991qx,Babu:1988mw,Babu:1989rb,Craig:2020bnv}), and by clever model-building explain how $\delta_{\rm \scriptscriptstyle CKM}$ is generated while $\bar \theta$ is not. These models have lots of interesting phenomenology and observational signatures. They also require many new colored fermions (sometimes introduced in very particular structures) and they often have the danger of being destabilized at loop order (see e.g. \cite{Dine:2015jga,Valenti:2021rdu,Vecchi:2014hpa,Valenti:2022uii,deVries:2021pzl,McNamara:2022lrw,Asadi:2022vys} for various concerns). 

An alternative approach takes advantage of $\theta$ being quite special as a CP-odd spurion, in that it can also be a spurion for anomalous chiral symmetries.
The relative benefit of such strategies is that they do not need to impose a discrete spacetime symmetry at any energy scale. 
Rather the structure of the theory is such that one particular CP-violating parameter, the strong CP angle, is naturally set to zero regardless of the CP violation present elsewhere in the theory, and we turn presently to explaining this structure.

Other approaches to strong CP can be found in, for example, \cite{Kuchimanchi:1995rp,Mohapatra:1995xd,Hiller:2001qg,Hiller:2002um,Hook:2014cda,Albaid:2015axa,Kaloper:2017fsa,Carena:2019nnd,Cherchiglia:2019gll,Dunsky:2019api,Valenti:2021xjp,Wang:2022osr,Dvali:2022fdv}, among many. Useful reviews include \cite{Cheng:1987gp,Dine:2000cj,Hook:2018dlk,Reece:2023czb}. 

\subsection{Peccei-Quinn for Strong CP}

The sort of solution to the strong CP problem which we will pursue engages directly with the fact that the theta angle can have an intricate spurionic structure in a theory of colored fermions, such that CP is not the only symmetry at play.

Consider an $SU(3)_C$ gauge theory which includes some $\theta$ topological angle,\footnote{We will often use the language of differential forms to emphasize when quantities are topological, and we remind the reader $\text{Tr} \left(G \wedge G\right) = \frac{1}{4} G^{\mu \nu} \tilde{G}_{\mu \nu}$, with $\tilde{G}^{\mu\nu} = \frac{1}{2} \epsilon^{\mu\nu\rho\sigma} G_{\rho\sigma}$.}
\begin{equation}
    S \supset i \frac{\theta}{8\pi^2}  \int_{\mathcal{M}} \text{Tr} \left(G \wedge G\right),
\end{equation}
with $G$ the strong gauge field strength. Obviously $\theta$ is an odd spurion for P or CP because the Levi-Civita symbol is a pseudotensor, but this is not necessarily the only spurionic charge for $\theta$.
We consider adding to our theory some colored fermions denoted generically $\lbrace q_i \rbrace$. Then for any global $U(1)_j$ symmetry rotation by an angle $\alpha_j$ acting on the fermions as
\begin{equation}
    U(1)_j: q_i \rightarrow q_i e^{i g_{ij} \alpha_j},
\end{equation}
for some integer charges $g_{ij}$, if this has a mixed anomaly,
\begin{equation}
    \mathcal{A}_{SU(3)^2_C \times U(1)_j}\equiv \mathcal{A}_j = \sum I_{i} g_{ij},
\end{equation}
where $I_i$ is the Dirac index of fermion $q_i$, then such a rotation of the quarks does not leave the theory invariant. Rather it effects a change in the partition function as
\begin{equation}
    Z \rightarrow Z \; \exp{\left( i \mathcal{A}_j \frac{\alpha_j}{8
    \pi^2} \int_{\mathcal{M}} \text{Tr} \left(G \wedge G\right) \right)}.
\end{equation}
This transformation effectively modifies the $\theta$ angle, or in other words $\theta$ has become a spurion transforming non-linearly under $U(1)_j: \theta \rightarrow \theta + \mathcal{A}_j \alpha_j$. In the quark sector such anomalous symmetries are often known as `Peccei-Quinn symmetries' \cite{Peccei:1977hh,Peccei:1977ur}. 
If $U(1)_j$ is a good symmetry, then the partition function remains invariant under an $\alpha_j$ transformation and so physics does not depend on the value of $\theta$. 

Now the effect of the anomaly is not just to give $\theta$ a spurionic charge, and the true structure here is a bit subtle. If there is an anomaly, $\mathcal{A}_j \neq 0$, then $U(1)_j$ is never truly a good, invertible symmetry---the ABJ anomaly from non-abelian instantons always implies explicit breaking of the order $\propto \exp(- 8 \pi^2/g^2)$. However, in an asymptotically free gauge theory $g^2 \rightarrow 0$ at high energies and the instanton violation decouples quickly, so it makes sense to talk about $U(1)_j$ as good symmetries in this limit. Then as one evolves to low energies, explicit violation of $U(1)_j$ appears from the dynamical growth of the gauge coupling. Solutions to the strong CP problem involving PQ symmetries make intrinsic use of this low-energy violation of what began as a good UV symmetry. 

The axion approach is to introduce a complex scalar $\phi$ which is charged under a $U(1)_{\rm \scriptscriptstyle PQ}$ and spontaneously breaks it. When $\phi$ gets a vev, its angular degree of freedom $a$ is a Goldstone for $U(1)_{\rm \scriptscriptstyle PQ}$, and so shifts nonlinearly under PQ symmetry transformations just as does $\theta$. 
In pure QCD, it is known quite generally that the gauge effects produce a potential $V(\theta)$ minimized at $\theta = 0$ \cite{Vafa:1984xg}. 
Now including an axion modifies the gluon action schematically as $i \int (\theta + a) G \tilde{G}$, and so QCD instantons generate a PQ-violating, but CP-preserving potential $V(a + \theta)$ which is minimized at $a = -\theta$. And while in pure QCD $V(\theta)$ just tells us about the relative vacuum energy of different theories with different $\theta$ parameter values, $a$ is now a dynamical degree of freedom. 

If $U(1)_{\rm \scriptscriptstyle PQ}$ began as a good symmetry, then this instanton effect generates the only potential for the axion. Then $a$ can cosmologically relax to this solution to `screen' whatever CP-violating $\theta$ angle was present and set the observed strong CP violation to vanish. 
Unfortunately the minimal version of this model which directly coupled $\phi$ to the SM quarks and so spontaneously broke one of the SM PQ symmetries, the Weinberg-Wilczek axion \cite{Weinberg:1977ma,Wilczek:1977pj}, has long since been ruled out. 
Axion models are revived by adding new, vector-like, colored fermions with their own $U(1)_{\rm \scriptscriptstyle PQ}$ symmetries which $\phi$ can spontaneously break. 
These `invisible axion' models have been a subject of increased interest in recent years, so much that we barely need mention they have rich, fascinating phenomenological signatures. But axions are not the only way to take advantage of a good $U(1)_{\rm \scriptscriptstyle PQ}$ to solve the strong CP problem.

The idea of the massless up quark solution was to instead posit that a SM $U(1)_{\rm \scriptscriptstyle PQ}$ is \textit{not} spontaneously broken. Since the up quark has the lightest mass in the infrared, one imagines an ultraviolet symmetry acting on the up quark in such a way as to forbid a mass term
\begin{equation}
    U(1)_{\rm \scriptscriptstyle PQ}: \bar u_1 \rightarrow \bar u_1 e^{i\alpha} \Rightarrow \det y_u = 0.
\end{equation}
A good such symmetry implies that strong CP violation vanishes, but there's a slight subtlety in how we talk about this given the standard definition of $\bar \theta$ in \eqref{eqn:bartheta}. Let us define the complex parameter $M \in \mathbb{C}$, which is the field-redefinition-invariant combination of the quark yukawa eigenvalues,
\begin{equation}
    M \equiv e^{i\theta} \det(y_u y_d),
\end{equation}
where the definition of the strong CP phase above is simply $\bar \theta \equiv \arg M$. The confusion is that this definition fails when $|M| \rightarrow 0$, such as when $U(1)_{\rm \scriptscriptstyle PQ}$ is imposed. Of course when the magnitude vanishes the phase is undefined, and this is sometimes discussed as `$\bar \theta$ becoming unphysical' in such a scenario. But this language merely encodes an artifact of using polar coordinates to parametrize $M$. Alternatively, we could work in Cartesian coordinates and then we have the CP-odd spurion 
\begin{equation}\label{masscp}
    \text{CP}: \text{Im}(M) \rightarrow - \text{Im}(M),
\end{equation}
which manifestly behaves smoothly as $|M| \rightarrow 0$. Clearly CP is preserved when $\text{Im}(M) = 0$, and $\bar \theta = 0$ corresponds to $M \in \mathbb{R}_+$. When we have a good $U(1)_{\rm \scriptscriptstyle PQ}$ symmetry, $|M| \rightarrow 0$ and we could say `$\bar \theta$ is unphysical' or we can just say that the CP-odd mass parameter vanishes $\text{Im}(M) = 0$. 

That subtlety discussed, let us return to the massless up quark solution, having begun in the UV with such a $U(1)_{\rm \scriptscriptstyle PQ}$ symmetry. The obvious issue is that we have long known from current algebra and hadron masses that the up quark mass does not vanish in the infrared---$U(1)_{\rm \scriptscriptstyle PQ}$ has been broken.
But this is sensible, since it was an anomalous symmetry. Just as QCD instantons violating $U(1)_{\rm \scriptscriptstyle PQ}$ provide a potential for the axion which localizes $\bar \theta$ to the CP-conserving value, so too could they potentially accord a good UV PQ symmetry with the observed quark masses.
Indeed the contributions of instantons to the masses of quarks automatically preserve the form of the CKM matrix in which $\bar \theta = 0$ \cite{Georgi:128245,Kaplan:1986ru,Choi:1988sy,Srednicki:2005wc}. 
That is to say the instanton effects will violate $|M|=0$ but only along the real axis. Whatever phases appear in the yukawas, we continue to have $\text{Im}(M)=0$ as a basis-independent statement.
The idea is then that $m_u$ might be zero in the UV, corresponding to a high quality $U(1)_{\rm \scriptscriptstyle PQ}$, which is broken solely by QCD instantons that provide the observed $m_u > 0$. 

After the massless up quark solution was proposed in the mid-80s, the observational status of this idea was held in limbo for some decades because the analytic calculation of instanton effects in QCD is not under theoretical control \cite{Banks:1994yg,Dine:2014dga}. 
Eventually, numerical computations of QCD on the lattice became powerful enough to resolve whether a vanishing up quark mass could fit data. Alas, the Standard Model does not bear out the massless up quark solution \cite{FlavourLatticeAveragingGroup:2019iem,Alexandrou:2020bkd}. In other words, one must begin at energies $\Lambda \gg \Lambda_{\rm \scriptscriptstyle QCD}$ with some non-trivial yukawa for the up quark already present in order to fit the far-infrared observables.

In recent years it has been realized that this solution may be revived in UV completions of the SM in which $SU(3)_C$ emerges from a larger gauge group in which it is non-trivially embedded \cite{Agrawal:2017evu,Agrawal:2017ksf} (see also earlier work on the possible relevance of small color instanton effects e.g. \cite{Holdom:1982ex,Flynn:1987rs,Choi:1998ep}). 
In some cases, instantons from the UV scale of this breaking $H \rightarrow SU(3)_C$ can provide a dominant contribution to the mass of the up quark (or the axion potential \cite{Csaki:1998vv,Csaki:2019vte}). 
This possibility has been studied in flavor deconstruction \cite{Agrawal:2017evu} and in composite Higgs models \cite{Gupta:2020vxb}.

Here we identify extensions of the SM with gauged quark flavor symmetries in which $\theta$ becomes a spurion for a non-invertible Peccei-Quinn symmetry. This understanding of the generalized symmetry structure of the Standard Model reveals a minimal realization of a small instanton approach to the massless quark solution in the context of color-flavor unification. 
That such non-trivial structures are even possible with the SM chiral matter content we find very suggestive. 
And this approach will have various benefits we will see below, including the possibility to separate the scales of the instanton effects at a UV symmetry-breaking scale and the flavor-breaking effects at the scale $H \rightarrow SU(3)_C$.
By starting with more of the SM's approximate global symmetries as gauge symmetries, we begin with an extremely simple theory and must understand how the SM structure is generated.
But unification has tremendous reductionist appeal, and the grand challenge in this model will be no more and no less than understanding a fully predictive theory of the quark yukawas. In this work we will factorize issues and show how this new strong CP solution will work whether or not we get the yukawas exactly correct, which we will leave to a future pursuit.

\section{Massless Quarks from Non-Invertible Peccei-Quinn Symmetries} \label{sec:fromIR}

We seek to generate yukawa couplings by the breaking of non-invertible chiral symmetries in certain extensions of the SM which result from gauging an (approximate) global symmetry of the SM fields. In the lepton sector, this analysis lead to a model of neutrino masses \cite{Cordova:2022fhg}. In the quark sector, such chiral symmetries might be termed `non-invertible Peccei-Quinn symmetries', and we investigate them by examining anomalies of SM fields. The material we present and our setup are aimed toward the strong CP problem, but our analysis of non-invertible symmetries may potentially be of broader interest. 

The analysis of this section is brief and self-contained.  A deeper analysis of non-invertible chiral symmetry can be found in \cite{Choi:2022jqy, Cordova:2022ieu}. We also provide a review of fractional instantons associated with non-trivial global structure in Appendix~\ref{app:fractionalInstanton} as well as a discussion on the global structure of the SM gauge group in Appendix~\ref{app:globalStructure}.\footnote{The technique employed in the appendices follows that of \cite{Anber:2021iip} which uses equations called `cocycle conditions' to systematically determine the most general possible global structure of the gauge group and spectrum of fractional instantons including those utilizing abelain factors of the gauge group. These configurations are called 'Color-Flavor-$U(1)$' (CFU) instantons in \cite{Anber:2021iip}.}

\subsection{Non-Invertible Chiral Symmetry}
\label{sec:nis}

To start, let us briefly recall some aspects of non-invertible chiral symmetry \cite{Choi:2022jqy, Cordova:2022ieu} that will appear below.  These symmetries often arise in chiral gauge theories.  The simplest example concerns a current $J_{\mu}$ for a chiral symmetry which is violated by an ABJ anomaly:
\begin{equation}\label{anomeqn}
    \partial^{\mu}J_{\mu}=\frac{k}{32\pi^{2}}F_{\alpha \beta}F_{\gamma \delta}\varepsilon^{\alpha \beta \gamma \delta},
\end{equation}
where in the above $k\in \mathbb{Z}$ is an integral anomaly coefficient, and $F_{\alpha\beta}$ is the field strength of an abelian gauge field.  The current $J_{\mu}$ is no longer conserved, but as is well-known, the absence of abelian instantons implies that at the level of the S-matrix (or relatedly local operator correlation functions) the chiral symmetry selection rules are still enforced.  Non-invertible symmetry gives a way to understand this phenomena non-perturbatively and to understand the unique features of such chiral symmetries.  The key idea is to recognize the right-hand side of \eqref{anomeqn} is a composite operator which is itself built from symmetry currents.  Indeed, the Bianchi identity implies that the field strength operator is closed, or more explicitly:
\begin{equation}\label{onefcons}
    \partial^{\mu}\left(\varepsilon_{\alpha \beta \mu \nu}F^{\alpha \beta}\right)=0~.
\end{equation}
This signals the presence of a higher symmetry: the magnetic one-form symmetry of an abelian gauge theory \cite{Gaiotto:2014kfa}.  The charged objects under this symmetry are 't Hooft lines, which physically represent the worldlines of probe magnetic monopoles.    

The appearance of a higher-form symmetry current in the anomaly equation \eqref{anomeqn} enables us to construct the operator (symmetry defect) that performs finite chiral symmetry transformations on Hilbert space.  Let $\Sigma$ denote the spatial slice (fixed time locus) where we wish to transform the fields by a chiral rotation by a finite angle $2\pi/kN$ for integral $N$. 
If the symmetry were invertible ($k=0$), then we would simply construct a symmetry defect operator for an arbitrary phase $\alpha$ as
\begin{equation}
    U_{\alpha}[\Sigma]= e^{i\alpha \int_{\Sigma} J},
\end{equation}
where we recognize $Q[\Sigma] \equiv \int_{\Sigma} J$ as the Noether Charge. With the ABJ anomaly this fails to be topological, and a new ingredient is needed. In addition to the exponentiated integrated current, along $\Sigma$ we must also include new topological degrees of freedom to cancel the ABJ anomaly and define a consistent conserved charge.  These topological degrees of freedom are charged under a one-form symmetry and hence can couple to the bulk through the magnetic one-form symmetry current discussed above.  In more detail, the Lagrangian on $\Sigma$ for the additional topological degrees of freedom involves a new dynamical $U(1)$ gauge field $C_{\mu}$ and takes the form of a Chern-Simons action:
\begin{equation}
    \mathcal{L}_{\Sigma}=\frac{iN}{4k}\int_{\Sigma}d^{3}x C_{\mu}\partial_{\nu}C_{\sigma}\varepsilon^{\mu\nu\sigma}+\frac{i}{2\pi}\int_{\Sigma}d^{3}x C_{\mu}\partial_{\nu}A_{\sigma}\varepsilon^{\mu\nu\sigma},
\end{equation}
where $F^{\mu\nu} = \partial^\mu A^\nu - \partial^\nu A^\mu$.
More generally, this construction can be carried out for any finite chiral transformation by a rational angle, resulting in a conserved charge (topological operator) that can implement the desired chiral transformation.  The consequence of coupling the additional topological degrees of freedom on the worldvolume means, in modern terminology that the symmetry has become non-invertible.  In other words it is no longer represented by unitary operators acting on Hilbert space.

The non-invertible nature of the chiral symmetry transformation has important technical and physical consequences.  To illustrate these, let us denote by $\mathcal{D}_{kN}(\Sigma)$ the operator on $\Sigma$ described above which implements a finite chiral transformation and let $\overline{\mathcal{D}}_{kN}(\Sigma)$ denote the transformation by the opposite angle.  When composed these transformations do not equal unity, but instead leave behind a condensate of magnetic one-form symmetry operators \cite{Choi:2021kmx, Kaidi:2021xfk,Roumpedakis:2022aik,Choi:2022jqy, Cordova:2022ieu}:
\begin{equation}\label{fusion}
    \mathcal{D}_{kN}(\Sigma)\times \overline{\mathcal{D}}_{kN}(\Sigma) \sim \sum_{\text{two-cycles}~ S\subset \Sigma}\exp\left(\frac{ i}{16\pi N}\int_{S}F_{\alpha \beta}dS_{\mu\nu}\varepsilon^{\alpha\beta \mu\nu}\right).
\end{equation}
One term on the right-hand side, corresponding to a trivial cycle $S$, is the unit operator, which is the naive result of the multiplication.  This is the only relevant term when acting on local operators.  The remaining terms on the right-hand side are visible only when acting on states carrying magnetic charge, or equivalently on 't Hooft line operators. 

The interplay of magnetic charge and the chiral symmetry transformation is the inevitable result of the anomaly \eqref{anomeqn}.  Indeed, while there are no abelian instantons in spacetimes with trivial topology, in a richer geometry such as those produced effectively by magnetic charges, abelian instantons are possible and the anomaly can be saturated.  More formally, it is natural to view spacetime $\mathbb{R}^{4}$ with an infrared regulator as topologically a four-sphere $S^{4}$.  The absence of abelian instantons is then a consequence of the fact that there are no topologically non-trivial two-cycles and hence no locus where a magnetic flux can be non-trivial.  By contrast, including an 't Hooft line effectively modifies the spacetime topology: the two-sphere linking the worldline of the charge is now a non-trivial two-cycle and in general in such configurations abelian instantons exist.  Thus, the non-invertible symmetry analysis above gives a formal way to understand selection rules that arise from the absence of instantons on $S^{4}$.  Note that this also reveals the importance of the fact that the right-hand side of \eqref{anomeqn} is composed of abelian field strengths.  In general for non-abelian field strengths instanton configurations exist already on $S^{4}$ and hence there are no-resulting non-invertible chiral symmetries. 

A crucial physical consequence of the fusion algebra \eqref{fusion} is that it reveals a channel for non-invertible symmetry breaking.  Consider a model with dynamical magnetic monopoles.  Below the scale of their mass, the worldline of a monopole appears in the infrared effective abelian gauge theory as a 't Hooft line.  Above the higgsing scale, the monopoles reveal themselves as fluctuating modes and the magnetic one-form symmetry is broken \cite{Cordova:2022rer}.  In other words, at this scale \eqref{onefcons} no longer holds.  However, since the magnetic one-form charges appear in the algebra of the chiral symmetries \eqref{fusion}, these in turn must also be broken by the presence of magnetic charges.  At a practical level, this means that loops of magnetic monopoles break the chiral symmetry.  Since magnetic monopoles are solitons the effects of such loops are generally non-perturbative in gauge theory couplings. To estimate their size, parameterize their mass as
\begin{equation}
    m_\text{mon}\sim \frac{v}{g}, 
\end{equation}
where $v$ is the Higgs vev, and $g$ the gauge coupling.  A monopole loop exists for a characteristic proper $\delta \tau$ that is inversely proportional to the effective cutoff set by the W-boson mass and hence $\delta \tau \sim 1/(gv).$  The corrections from monopole loops therefore have a characteristic size:
\begin{equation}\label{dlis}
    \delta \mathcal{L}\sim \exp\left(-S_{\text{mon}}\right) \sim \exp\left(-m_{\text{mon}}\delta \tau\right)\sim \exp\left(-\#/g^{2}\right).
\end{equation}
Thus, when the chiral symmetry violation is mediated by one-form symmetry breaking effects of magnetic monopoles, the above estimate yields the expected size of the leading corrections.  

We also note that \eqref{dlis} is precisely the characteristic size of non-abelian instanton effects.  Indeed loops of magnetic monopoles and more generally dyons, can be viewed as describing instanton corrections to the action \cite{Fan:2021ntg}.  This is natural since a dyon has non-vanishing $E\cdot B$ and hence can saturate the topological term in the action.  The breaking of a non-invertible chiral symmetry by magnetic monopole loops can therefore alternatively be understood as the direct breaking of the chiral symmetry in the non-abelian gauge theory by instanton effects.  Turned around, the lens of non-invertible symmetry provides a key tool to understand when instantons will yield the leading contribution to a given physical process, and can thus guide our hand towards interesting UV models.  This is the model building perspective that we will utilize below.  

Thus far, we have seen how non-invertible chiral symmetry provides a language for dealing with abelian ABJ anomalies.  In our application below we will instead require a more sophisticated version of this construction which goes beyond the physics of abelian gauge theories.  Specifically we will employ non-invertible chiral symmetries that exist due to the presence of discrete magnetic one-form symmetries in non-abelian gauge theories.  

In general, discrete magnetic one-form symmetries occur in non-abelian gauge theories where the gauge group has a non-trivial fundamental group, which we often refer to as a non-trivial global structure.  For instance, the gauge group $SU(N)$ does not have any discrete magnetic one-form symmetry, while the gauge group $SU(N)/\mathbb{Z}_{N}$ has magnetic symmetry $\mathbb{Z}_{N}.$  Physically, this means that the gauge group $SU(N)/\mathbb{Z}_{N}$ has a richer spectrum of magnetic monopoles, (realized as non dynamical 't Hooft lines) which carry conserved $\mathbb{Z}_{N}$ quantum numbers.  Crucial for our purposes, discrete magnetic one-form symmetry also signals the presence of fractional instantons.  The hallmark of such field configurations is that the instanton number is no longer an integer, but instead its fractional part can be expressed as a suitable square of a discrete magnetic flux (see e.g. \eqref{infracfirst} below). As above, this implies that in a topologically trivial setup, without two-cycles or equivalently 't Hooft lines, the minimal allowed instanton number cannot be saturated.  Indeed, on $\mathbb{R}^{4}$ (or more precisely the IR regulated $S^{4}$) non-abelian instanton numbers are always integral.

In the context of ABJ anomalies and chiral symmetries, fractional instantons therefore have an effect similar to the abelian instantons discussed above.  In particular, there can again be chiral symmetries where the minimum allowed anomaly coefficient cannot be saturated on $S^{4}$.  In this situation the, typically discrete, chiral symmetry is then in fact non-invertible.  The symmetry defect operator supports a fractional hall state which now couples to the bulk non-abelian gauge fields through the discrete magnetic flux.  (For more explicit formulae see \cite{Cordova:2022ieu}.)
Thus, discrete non-invertible chiral symmetries depend in detail on the precise global form of the gauge group and can in turn point towards specific UV completions where these symmetries are in turn violated by new loops of dynamical monopoles, or equivalently new small instanton effects.  We exploit these mechanisms in our models below. 

\subsection{Approximate Symmetries of the Standard Model}

We now turn to apply these considerations in the SM and beyond.  We recall the basic structure of the SM fermions as reviewed in Table \ref{tab:charges}.  Throughout we work in conventions where hypercharge is integrally normalized.
{\setlength{\tabcolsep}{0.6 em}
\renewcommand{\arraystretch}{1.3}
\begin{table}[h]\centering
\large
\begin{tabular}{|c|c|c|c|c|c|c|c|}  \hline
 & $  Q_i  $ & $ \bar u_i $ & $ \bar d_i $ & $  L_i  $ & $ \bar e_i $ & $  N_i  $ & $ H $ \\ \hline
$SU(3)_C$ & $\mathbf{3}$ & $\bar{\mathbf{3}}$ & $\bar{\mathbf{3}}$ & -- & -- & -- & --\\ \hline
$SU(2)_L$ & $\mathbf{2}$ & -- & -- & $\mathbf{2}$ & -- & -- & $\mathbf{2}$\\ \hline
$U(1)_{Y}$ & $+1$ & $-4$ & $+2$ & $-3$ & $+6$ & -- & $-3$\\ \hline
$U(1)_{B}$ & $+1$ & $-1$ & $-1$ & -- & -- & -- & --\\ \hline
$U(1)_{L}$ & -- & -- & -- & $+1$ & $-1$ & $-1$ & --\\ \hline 
\end{tabular}\caption{Representations of the Standard Model Weyl fermions under the classical gauge and global symmetries. We normalize each $U(1)$ so the least-charged particle has unit charge. We list also the charges of the right-handed neutrino $N$ and the Higgs boson $H$.  }\label{tab:charges}
\end{table}}

The SM includes the yukawa interactions coupling the fermions to the Higgs field $H$:
\begin{equation}\label{eqn:SMyuk}
    \mathcal{L} \supset y_u^{ij} \tilde{H} Q_i \bar u_j + y_d^{ij} H Q_i \bar d_j + y_e^{ij} H L_i \bar e_j~,
\end{equation}
where $\tilde{H} \equiv i \sigma_2 H^\star$.  The observed yukawa matrices $y$ (equivalently, the fermion masses and flavor changing processes) explicitly break all of the non-abelian continuous global symmetries, as they provide different masses for the generations. We will consider models which, to start, have vanishing down-type yukawas $y_{d}\rightarrow 0$ and later aim to regenerate these couplings through symmetry breaking effects.  Our  analysis will reveal that the down-type quark yukawa $y_d$ can be protected by certain kinds of non-invertible symmetries. We note that the same analysis with instead $y_u \leftrightarrow y_d$ would give the same conclusions, but this will ultimately be the right choice for our purposes. At the classical level, non-vanishing, general $y_u$ leaves the following abelian flavor symmetries unbroken: 
\begin{equation}
   \prod_i^3 U(1)_{\tilde{B}_i} \times U(1)_{\bar d_i}, 
   \label{eq:classical_symm}
\end{equation}
where $\tilde{B}_i$ and the conventional baryon number $B_i$  are defined as:
\begin{equation}
    \tilde{B}_i = Q_i - \bar{u}_i, ~~~B_i=\tilde{B}_i - \bar{d}_i=Q_i - \bar{u}_i-\bar{d}_{i}.
\end{equation}
In fact the full $U(3)_{\bar d}$ is thus far a full symmetry, but it is the analysis of the $U(1)$ factors which is pertinent below.

In Table~\ref{tab:IRanomaly} we list the Adler-Bell-Jackiw \cite{Adler:1969gk,Bell:1969ts} anomaly coefficients of these global symmetries with the SM gauge group and additional gauge groups appearing in extensions discussed below. 
One immediate lesson we can extract from Table~\ref{tab:IRanomaly} is that there exists no non-invertible symmetry within the quark sector with gauge group $SU(3)_C \times SU(2)_L \times U(1)_Y / \Gamma$ with $\Gamma = 1$. This is because each of these $U(1)$ global symmetries and any linear combinations of them are dominantly broken by non-abelian instantons, or they remain as good invertible symmetries, e.g.~$\bar d_1 - \bar d_2$. Indeed, as reviewed in section \ref{sec:nis} above, non-invertible chiral symmetry arises from a classical $U(1)$ dominantly broken either by an abelian instanton or a fractional instanton of non-abelian gauge theory. As we show in Appendix~\ref{app:no_NIS_SM}, the absence of non-invertible symmetries of the quark sector with up yukawas turned on remains the case even with non-trivial global structure $\Gamma \in \lbrace \mathbb{Z}_6, \mathbb{Z}_3, \mathbb{Z}_2\rbrace$. 

\begin{table}[t]\centering
\large
\begin{tabular}{|c|c|c|c|c|c|c|}  \hline
 & $U(1)_{\tilde{B}_1}$  & $U(1)_{\tilde{B}_2}$&$U(1)_{\tilde{B}_3}$ & $U(1)_{\bar d_1}$  & $U(1)_{\bar d_2}$&$U(1)_{\bar d_3}$ \\ \hline
 
$SU(3)_C^{2}$ & $+1$ & $+1$ & $+1$ & $+1$ & $+1$ & $+1$\\ \hline 

$SU(2)_L^{2}$ & $N_c$ & $N_c$ & $N_c$ & $0$ & $0$ & $0$\\ \hline 

$U(1)_Y^{2}$ & $-14 N_c$ & $-14 N_c$ & $-14 N_c$ & $4 N_c$ & $4 N_c$ & $4 N_c$\\ \hline 

$U(1)_{H}^{2}$ & $N_c$ & $N_c$ & $4 N_c$ & $N_c$ & $N_c$ & $4 N_c$\\ \hline 

$U(1)_{Y}U(1)_{H}$ & $-2N_c$ & $-2N_c$ & $4N_{c}$ & $-2N_c$ & $-2N_c$ & $4 N_c$\\ \hline 

$[\text{CH}]$ & $1$ & $1$ & $2$ & $1$ & $1$ & $2$\\ \hline 

\end{tabular}\caption{ABJ anomalies of chiral symmetries of quark sector with SM gauge groups and a gauged $U(1)_{H} \equiv U(1)_{B_1 + B_2 - 2 B_3}$. We also show the anomaly coefficients in the fractional instanton background, denoted as $[\text{CH}]$ (for ``color-H'' admixture), appearing in the $SU(3)_C \times U(1)_H / \mathbb{Z}_3$ extension.}
\label{tab:IRanomaly}
\end{table}

So far then, our analysis has revealed no surprising symmetries.  With vanishing down quark yukawas some PQ symmetries are restored and hence strong CP is conserved.  
But such a scenario, like the massless up quark solution, is excluded since it fails to reproduce the observed IR physics of QCD.  Instead, we will consider two minimal extensions of the SM which enjoy non-invertible chiral symmetry acting on the quarks and suggest a natural UV completion. The new symmetries we find will protect the down yukawa couplings in a way that UV symmetry-breaking can generate the observed nonzero values while setting the strong CP phase $\bar{\theta}$ to zero. 

\begin{figure}
  \centering
  \includegraphics[width=.66\textwidth]{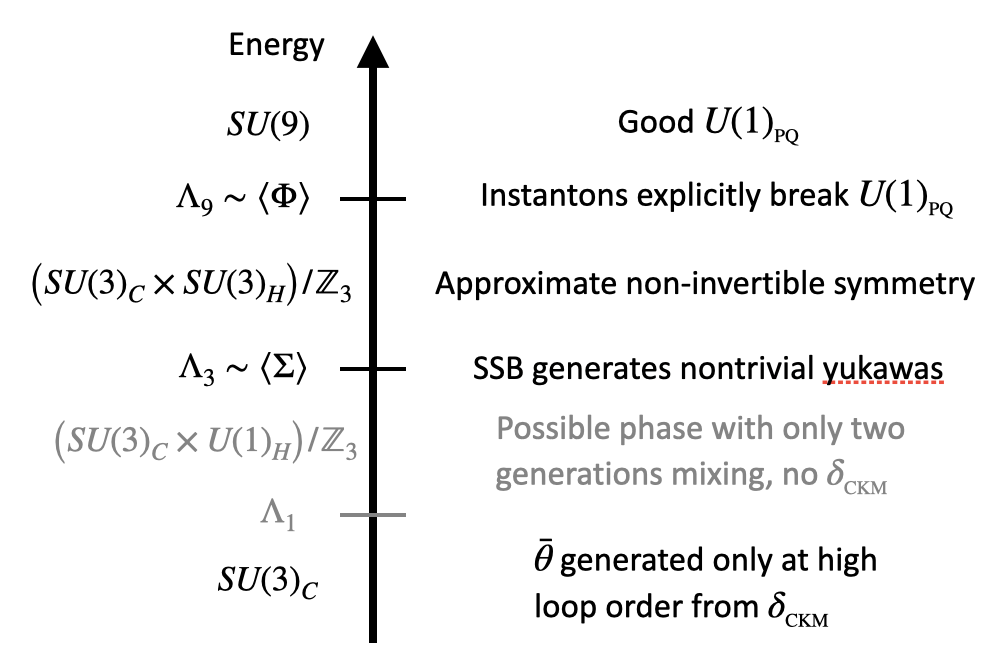}\hspace{.5cm}
  \caption{For visual reference, the organization of phases governing the quarks in this work. In the model of Section \ref{sec:fromUV} we do not make use of the possible $U(1)_H$ phase, though there is also a non-invertible symmetry in this phase as discussed in Section \ref{subsec:U1H_ext}.}
  \label{fig:RGFlow}
\end{figure}

Our extensions are based on gauging certain `horizontal' symmetries of the quark sector. In Section~\ref{subsec:U1H_ext}, we discuss a $Z'$ extension by a $U(1)_H$ gauge group with
\begin{equation}\label{his}
    H=B_1+B_2-2B_3,
\end{equation}
and an extension by non-abelian $SU(3)_H$ gauged horizontal flavor symmetry will be presented in Section~\ref{subsec:SU3H_ext}. Each model below may be viewed as a separate candidate extension of the SM.  Alternatively, it is also possible to think of them as two different phases of the same theory along a renormalization group flow. Starting from an UV theory (for instance the $SU(9)$ model presented in Section~\ref{sec:fromUV}), one flows to the $SU(3)_H$-extension as an intermediate phase. Further flowing to the IR leads to the $U(1)_H$-extension. Finally, spontaneous breaking of $U(1)_H$ at yet a lower scale brings us to the SM. (See Figure \ref{fig:RGFlow}.)  Ultimately, as discussed in Section~\ref{sec:fromUV}, to achieve the IR physics of the SM we must break the new chiral symmetries under consideration.  First however, we present the approximate symmetries from the infrared as a key guide to our model.  

\subsection{\texorpdfstring{$U(1)_H$}{U(1)H} Extension}
\label{subsec:U1H_ext}

One interesting possible $Z'$ extension which is free of cubic and mixed anomalies (hence can be gauged with no additional matter) is the horizontal baryon number symmetry $U(1)_H$ given by \eqref{his}.\footnote{We note also that this theory has an (approximate) higher group structure $SU(2)_{Q,u,d}^{(0)} \times_\kappa U(1)_H^{(1)}$, but it will not play an important role for our purposes (c.f. \cite{Cordova:2022qtz}).} Similar lepton family difference symmetries which are anomaly-free (e.g. $U(1)_{L_\mu - L_\tau}$) have received much attention as $Z'$ models for reasons both formal and phenomenological. These are exact symmetries of the Standard Model with zero neutrino masses, but they are not exact symmetries of the real world. The similarity of the lepton and quark sectors in the SM (especially if neutrinos are Dirac) suggests investigating also the quark family difference symmetries. Indeed, these are also approximate global symmetries of nature, which are less well-preserved in the infrared  because the quark yukawas are much larger. Further motivation for this particular choice of $U(1)_H$ will be given presently. 

Before proceeding with a detailed analysis of the symmetries, we note a crucial discrete identification of the gauge and global symmetry groups.  Specifically, with the matter content of the SM the following groups act identically on all fields  
\beq 
\left[ \mathbb{Z}_3^{H} \subset U(1)_H \right] = \left[ \mathbb{Z}_3^{C} \subset SU(3)_C \right] = \left[ \mathbb{Z}_3^{Y} \subset U(1)_Y \right] = \left[ \mathbb{Z}_3^{B} \subset U(1)_B \right].
\label{eq:Z_3_equivalence}
\eeq 
This means that various diagonal $\mathbb{Z}_{3}$ subgroups act trivially on all fields and we can modify the theory by quotienting the gauge group by any of these trivially acting subgroups. In our case, the relevant choice is between the gauge groups
\begin{equation}\label{globalHchoice}
    SU(3)_{C}\times U(1)_{H} ~~\text{vs.}~~\left(SU(3)_{C}\times U(1)_{H}\right)/\mathbb{Z}_{3},
\end{equation}
while other possible quotients do not play a role in our analysis.  This choice can be clarified physically in the language of higher symmetries as follows.  If we do not quotient by $\mathbb{Z}_{3}$, then the theory has an electric $\mathbb{Z}_{3}$ one-form global symmetry whose charged objects are Wilson lines that cannot be screened by dynamical matter.  By contrast in the theory with the $\mathbb{Z}_{3}$ quotient, these representations are removed and correspondingly Dirac quantization allows new magnetic monopoles whose 't Hooft lines are charged under a dual $\mathbb{Z}_{3}$ magnetic one-form symmetry.  Relatedly, the theory with the quotient admits fractional instantons. These effects will be crucial to our theory and hence we focus on the case where the gauge group has the $\mathbb{Z}_{3}$ quotient (right-hand side of \eqref{globalHchoice}).  We note that this possible gauge group global structure relies precisely on our choice of horizontal gauge group, and exists due to the fact that $N_{g}=N_{c}.$

The ABJ anomaly coefficients of the global symmetries listed in \eqref{eq:classical_symm} with $U(1)_H$ are given in Table~\ref{tab:IRanomaly}. In searching for non-invertible symmetries, we are interested in global $U(1)$ factors which are anomalous only due to $U(1)_H$ effect (or at least where $U(1)_H$ provides the dominant anomalous breaking). To guide us it is useful to notice that SM anomalies are flavor-universal and therefore $\tilde{B}_i - \tilde{B}_j$, and $\bar d_i - \bar d_j$ with any choice of $i,j=1,2,3$ are SM-anomaly-free. There are then four obvious candidates:
\beq \label{adefs}
A_1 = \bar{d}_3 - \bar{d}_1, \;\; A_2 = \bar{d}_3 - \bar{d}_2, \;\; A_3 = \tilde{B}_3- \tilde{B}_1, \;\; A_4 = \tilde{B}_3 - \tilde{B}_2.
\eeq 
However we note that the combination
\begin{equation}
    A_{1}+A_{2}-A_{3}-A_{4}=H,
\end{equation}
is gauged so the generators in \eqref{adefs} above represent three possible flavor symmetries.

Consider first the effect of the familiar ABJ anomaly on these symmetries.  From Table \ref{tab:IRanomaly} we have the anomaly coefficients:
\begin{equation}
    \left[ U(1)_{A_{i=1,\cdots, 4}} \right] \left[ U(1)_H \right]^2 = 3 N_c, ~~~  \left[ U(1)_{A_{i=1,\cdots, 4}} \right] \left[ U(1)_Y \right]\left[ U(1)_H \right]=6N_{c}.
\end{equation}
This breaks each $U(1)_{A_{i}}$ to a $\mathbb{Z}_{3N_c}$ invertible symmetry.  Meanwhile, the rest of $U(1)_{A_{i}}$ forms an infinite non-invertible symmetry acting as rotations with rational angle different from $\mathbb{Z}_{3N_c}$.  In the case where the gauge group does not have a $\mathbb{Z}_{3}$ quotient, this concludes the analysis and the down yukawas are spurions of invertible symmetries. However in the case with $\left(SU(3)_{C}\times U(1)_{H}\right)/\mathbb{Z}_{3}$ additional fractional instantons further modify these symmetries.

\subsubsection{Symmetry Breaking from Fractional \texorpdfstring{$\left(SU(3)_{C}\times U(1)_{H}\right)/\mathbb{Z}_{3}$}{} Instantons}

We now analyze the effects of fractional instantons on the symmetries.  Focusing on $U(1)_{H}/\mathbb{Z}_{3},$ the first observation is that  the magnetic flux is now fractionally quantized in units of one-third:
\begin{equation}\label{ffrac}
    \int_{\Sigma} \frac{F_{H}}{2\pi} \in \frac{1}{3}\mathbb{Z}.
\end{equation}
Those fluxes above which are integrally quantized are standard field configurations of $U(1)_{H},$ while those which are fractional are new configurations allowed by the quotient.   More subtly, the $SU(3)_{C}/\mathbb{Z}_{3}$ also admits new flux configurations that are not allowed in $SU(3)_{C}$.  These are labelled by a discrete analog of the magnetic flux, sometimes referred to as a second Stiefel-Whitney class:
\begin{equation}
\omega(A_{C})\in H^{2}(M,\mathbb{Z}_{3}),
\end{equation}
where above $M$ is the spacetime four manifold and $A_{C}$ is the color gauge field.  Concretely, this means that $\omega$ is an object which may be integrated over any two-cycle $\Sigma$ in spacetime yielding an integer which is well-defined modulo 3: 
\begin{equation}\label{w2is}
    \int_{\Sigma} \omega(A_{C}) \in \mathbb{Z}_{3},
\end{equation}
and should be viewed as the non-abelian analog of the fractional part of the magnetic flux in \eqref{ffrac}.  

In our situation, the gauge group $(SU(3)_{C}\times U(1)_{H})/\mathbb{Z}_{3}$ does not have independent quotients, but one quotient that acts simultaneously on the two factors.  This in turn implies that the fractional magnetic fluxes between the abelian and non-abelian factor are correlated which we express as:
\begin{equation}\label{fracfrac}
    \frac{F_{H}}{2\pi}=\frac{1}{3}\omega(A_{C})-X.
\end{equation}
Here  $X \in H^{2}(M,\mathbb{Z})$ can be viewed as a standard quantized flux, and each term thought of as a cohomology class (equality holds upon integration on any two-cycle). 

The final technical tool we need is an analysis of the instanton number for gauge groups with discrete quotients. As reviewed in Appendix~\ref{app:fractionalInstanton}, the quotient implies that the instanton number $\mathcal{N}_{C}$ of $SU(3)_{C}$ is no longer integral, but is in general fractional, with the fractional part controlled by the discrete magnetic flux \eqref{w2is}
\begin{equation}\label{infracfirst}
   \mathcal{N}_{C}= \frac{1}{8\pi^{2}}\int_{M}\mathrm{Tr}\left(F_{C}\wedge F_{C}\right)=\frac{1}{3}\int_{M} \omega \wedge \omega ~~\mod 1,
\end{equation}
where the notation above means that the fractional part of the left and right-hand sides agree.  Using \eqref{fracfrac} we can also express this in terms of the flux $F_{H}$: 
\begin{eqnarray}
    \mathcal{N}_{C} & = & 3\int_{M}\left(\frac{F_{H}}{2\pi}+X\right)\wedge\left(\frac{F_{H}}{2\pi}+X\right) ~~\mod 1 \nonumber\\
     & = & 6\left(\frac{1}{8\pi^{2}} \int_{M}F_{H}\wedge F_{H}\right)+6\int_{M}\frac{F}{2\pi}\wedge X+3 \int_{M}X \wedge X ~~\mod 1 \label{qcqh}\\
     & = & 6\left(\frac{1}{8\pi^{2}} \int_{M}F_{H}\wedge F_{H}\right) ~~\mod 1 \nonumber\\
     & = & 6\mathcal{N}_{H}~~ \mod 1,\nonumber
\end{eqnarray}
where in the third line we have used \eqref{fracfrac} to show that the contributions involving $X$ do not modify the fractional part, and in the last line we have introduced $\mathcal{N}_{H}$ which is the instanton density of $U(1)_{H}/\mathbb{Z}_{3}:$
\begin{equation}
    \mathcal{N}_{H}=\frac{1}{8\pi^{2}}\int_{M}F_{H}\wedge F_{H}.
\end{equation}
The final result of \eqref{qcqh} shows that instantons of $(SU(3)_{C}\times U(1)_{H})/\mathbb{Z}_{3}$ have correlated fractional part, i.e. $\mathcal{N}_{C}$ can only be fractional if $\mathcal{N}_{H}$ is also fractional.  

We can now use the above to compute the most refined anomaly coefficients in the presence of fractional instantons and discover the final fate of the symmetries in our problem.    Using the notation $\psi_i = \left\lbrace Q_i, \bar{u}_i, \bar{d}_i \right\rbrace$ to denote a general charged fermion, the Dirac index for $\psi_i$ is an instanton background is computed as
\beq  
I_{\psi_i} = n_{\psi_i} T_{\psi_i} \mathcal{N}_C + \text{dim}_{\psi_i} n_{\psi_i} \left(q^{H}_{\psi_i}\right)^2 \mathcal{N}_H,
\label{eq:Dirac index}
\eeq 
where $n_{\psi_i}$ is the multiplicity of $\psi_i$ including both flavor and $SU(2)_L$ gauge degrees of freedom. $T_{\psi_i}$ denotes the Dynkin index of $\psi_i$ under $SU(3)_C$ (which is $1$ for all $\psi_i$), $\text{dim}_{\psi_i}$ is the dimension of $SU(3)_C$ representation (which is $3$ for all $\psi_i$), and finally $q^{H}_{\psi_i}$ is the $U(1)_H$ charge. Importantly, even though the instanton numbers $\mathcal{N}_{C}$ and $\mathcal{N}_{H}$ are individually in general fractional, the indices above are always integral due to \eqref{qcqh} and the fact that the matter content is consistent with the $\mathbb{Z}_{3}$ quotient. The anomaly coefficient for an abelian flavor symmetry $f$ in the presence of a general instanton background is then given by the formula:
\begin{equation}
    \mathcal{A}_{f}=\sum_{\psi_{i}} q^{f}_{\psi_{i}}I_{\psi_i}=3\mathcal{N}_{H}\sum_{\psi_{i}} q^{f}_{\psi_{i}}n_{\psi_i}\left(2+\left(q^{H}_{\psi_i}\right)^2\right)+k\sum_{\psi_{i}} q^{f}_{\psi_{i}}n_{\psi_i},
\end{equation}
where we have used \eqref{qcqh} as well as the details of our fermion spectra discussed above to simplify the index formula, and $k=\mathcal{N}_{C}-6\mathcal{N}_{H}$ is an integer.  The strongest constraints now come from choosing the most fractional instanton possible, which from \eqref{ffrac} is $\mathcal{N}_{H}=1/9$ and $k=0$ (so that $\mathcal{N}_{C}=2/3$).  Carrying out the sum then leads to the anomaly coefficients summarized in the final row of Table \ref{tab:IRanomaly}.

From this analysis, one sees that for each of the symmetries $A_{i}$ defined in \eqref{adefs} we have
\begin{equation}\label{afracanoms}
    \left[ U(1)_{A_{i=1,\cdots, 4}} \right] [\text{CH}] = 1.
\end{equation}
Therefore the fractional instantons of $(SU(3)_{C}\times U(1)_{H})/\mathbb{Z}_{3}$ turn each $U(1)_{A_i}$ completely into non-invertible symmetries.  Of these a discrete $\mathbb{Z}_{3}$ subgroup is particularly notable.  Consider the following equality of charges modulo three: 
\begin{equation}\label{Z3niv}
    \tilde B + \bar{d} = H+A_{1}+A_{2},~ \ (\text{mod} \ 3)
\end{equation}
As $H$ is gauged, this means that the diagonal flavor combination $A_{1}+A_{2}$ generates a discrete flavor symmetry, $\mathbb{Z}_{3}^{\tilde{B}+\bar{d}},$  that acts in a generation independent way. According to our analysis above, $\mathbb{Z}_{3}^{\tilde{B}+\bar{d}}$ is a non-invertible symmetry.  As we will see below this discrete symmetry plays a key role in protecting quark masses.  

More generally, our calculations of anomaly coefficient also allow us to deduce the subgroup of invertible symmetries, i.e. those that do not participate any anomalies involving $H$.  A general charge $J$ can be expressed in terms of integers $\ell_{i}$ as
\begin{equation}
   J= \ell_{1}A_{1}+\ell_{2}A_{2}+\ell_{3}A_{3}+\ell_{4}A_{4}.
\end{equation}
From \eqref{afracanoms}, the condition that $J$ defines an invertible symmetry is then 
\begin{equation}
    \ell_{1}+\ell_{2}+\ell_{3}+\ell_{4}=0.
\end{equation}
Modding out by the $H$ gauge redundancy leaves a rank two invertible flavor symmetry. These symmetries are summarized in Table \ref{tab:globalHorizontal} below. 
{\setlength{\tabcolsep}{0.6 em}
\renewcommand{\arraystretch}{1.3}
\begin{table}[ht]\centering
\large
\begin{tabular}{|c|c|}  \hline
Gauged & $U(1)_{B_{1}+B_{2}-2B_{3}}$ \\ \hline
Invertible & $U(1)_{\tilde{B}_3 - d_3 - \tilde{B}_1 + d_1} \times U(1)_{\tilde{B}_3 - d_3 - \tilde{B}_1 + d_2}$ \\ \hline
Non-Invertible & $U(1)_{\bar{d}_1 + \bar{d}_2 - 2 \bar{d}_3} \supset \mathbb{Z}_3^{\tilde{B}+\bar{d}}$ \\ \hline
\end{tabular}\caption{Symmetries of the Standard Model with no down yukawas after gauging $U(1)_H$ with non-trivial global structure.}\label{tab:globalHorizontal}
\end{table}}

\subsubsection{Massless Down Quarks from Non-Invertible PQ Symmetry}
\label{sec:masslessd}

Having identified all symmetries of the theory, we now discuss how the spurion structure of the down quark yukawa terms. 
\beq 
\c{L}_{y_d} = y_d^{ij} H Q_i \bar{d}_j.
\eeq
Of course in our analysis above we by hand set $y_{d}^{ij}\rightarrow 0$.  Thus, we will now see which symmetries must inevitably be broken to regenerate non-zero yukawas.

To begin we must enforce, $U(1)_H$ gauge invariance:
\bea
&& U(1)_H\text{--allowed:} \;\;\;\;~ Q_1 d_1, Q_2 d_2, Q_3 d_3 \;\;\; \text{(diagonal)} \\
&& \hspace{3.05cm} ~Q_1 d_2, Q_2 d_1 \hspace{1.3cm} \text{(off-diagonal)} \nonumber \\
&& U(1)_H\text{--forbidden:} \;~ Q_1 d_3, Q_3 d_1, Q_2 d_3, Q_3 d_2
\eea
This result is a simple consequence of the disparity in $U(1)_H$ charge assignments between the first two and third generations, which clearly implies there is no mixing of the third generation with the first or second.

Since $U(1)_{A_{1,2}}$ acts only on $\bar{d}$, it is quick to see that all $U(1)_H$-invariant components listed above are forbidden by non-invertible Peccei-Quinn symmetries.
In particular, we note that all $U(1)_{H}$ allowed components of the yukawa matrix are forbidden by the $\mathbb{Z}_{3}^{\tilde{B}+\bar{d}}$ non-invertible discrete symmetry. 
Thus, even if all other symmetries are broken, in the $U(1)_{H}$ phase with $\mathbb{Z}_{3}^{\tilde{B}+\bar{d}}$ the down quark yukawa matrix must vansish.  Conversely, we may also view non-zero entries of the down yukawa matrix as spurions of the non-invertible symmetry.\footnote{For completeness, we also note that the $Q_1 d_2$ and $Q_2 d_1$ components of the down yukawa matrix are further protected by the two anomaly-free $U(1)$ symmetries.} 

As discussed around \eqref{masscp}, we see that when the non-invertible Peccei-Quinn symmetry is preserved there are massless quarks and CP is a symmetry of the strong sector of the SM.  For this reason, such models inform us about \emph{massless down-quark solutions} to the strong CP problem.  We note that this class of models, and in particular the tight interplay with non-invertible symmetry only occurs for the choice of global form of the gauge group $(SU(3)_{C}\times U(1)_{H})/\mathbb{Z}_{3}$ which in turn is only possible because $N_{c}=N_{g}.$  Of course, in reality, down quarks are massive, implying that the non-invertible symmetry must be explicitly broken. The completion of the massless down-quark solution, therefore, requires a mechanism of non-invertible symmetry violation and the generation of observed quark masses, mixings and the CKM phase. We discuss these in Section~\ref{sec:fromUV}.

\subsection{\texorpdfstring{$SU(3)_H$}{SU(3)H} Extension}
\label{subsec:SU3H_ext}

We now consider gauging the entire anomaly-free non-abelian quark flavor symmetry $SU(3)_H$.  We again do so with a non-trivial choice of quotient in the color-flavor gauge group:
\beq 
\frac{SU(3)_C \times SU(3)_H}{\mathbb{Z}_{3}}.
\label{eq:SU3H_global_structure_ourchoice}
\eeq 
The quarks transform in bi-fundamental representations  summarized in Table~\ref{tab:inter_charges}. For instance, the quark doublet (now bold-faced) $\mathbf{Q}$ is both a $3$ of $SU(3)_C$ unifying the color (red, green, blue) quantum numbers as well as a $3$ of $SU(3)_H$ unifying the flavor (e.g.~up, charm, top) quantum numbers as a gauge symmetry. That all colored SM matter obeys this pattern allows the non-trivial global structure chosen in \eqref{eq:SU3H_global_structure_ourchoice}.

\begin{table}\centering
\large
\renewcommand{\arraystretch}{1.3}
\begin{tabular}{|c|c|c|c|c|}  \hline
 & $SU(3)_C$ & $SU(3)_H$ & $U(1)_{B}$ & $U(1)_{\bar{d}}$ \\ \hline

$\mathbf{Q}$ & $3$ & $3$ & $+1$ & $0$ \\ \hline

$\mathbf{\bar u}$ & $\bar 3$ & $\bar 3$ & $-1$ & $0$ \\ \hline

$\mathbf{\bar d}$ & $\bar 3$ & $\bar 3$ & $-1$ & $+1$ \\ \hline

\end{tabular}\caption{Symmetry and matter content of the $SU(3)_H$-extension. }\label{tab:inter_charges}
\end{table}

Much of the symmetry analysis parallels that of the abelian horizontal extension discussed above.  We again consider the limit of vanishing down-type yukawas and further restrict the up-type yukawas to be family symmetric and hence compatible with $SU(3)_{H}$.  The relevant classical symmetries are now flavor independent:
\begin{equation}
\frac{U(1)_{B}}{\mathbb{Z}_{3}} \times U(1)_{\bar{d}},
\end{equation}
which have charges listed in Table~\ref{tab:inter_charges}.  Here, the $\mathbb{Z}_{3}$ quotient on $U(1)_{B}$ arises because that subgroup is gauged as noted in \eqref{eq:Z_3_equivalence}.

We can first consider the anomalies that do not probe the global structure of the gauge group.  The relevant anomaly coefficients are summarized in Table~\ref{tab:Interm_anomaly}.
The net breaking effect is given by the greatest common divisor (gcd) of all anomaly coefficients, and this shows that the classical global symmetries are broken down to
\beq
\frac{U(1)_{B}}{\mathbb{Z}_{3}} \times U(1)_{\bar{d}} \to \frac{\mathbb{Z}_9^{B}}{\mathbb{Z}_{3}} \times \mathbb{Z}_{3}^{\bar{d}}\cong \mathbb{Z}_3^{B} \times \mathbb{Z}_{3}^{\bar{d}}.
\eeq 

Without the quotient on the gauge group in \eqref{eq:SU3H_global_structure_ourchoice} the above concludes the analysis of the symmetries.  However with the $\mathbb{Z}_{3}$ quotient there are further instanton configurations to consider.   Specifically, there are now gauge fields with non-trivial Stiefel-Whitney classes for both the color and horizontal gauge group with:
\begin{equation}\label{frac3ext}
    w_{2}(A_{C})=w_{2}(A_{H})\in H^{2}(M,\mathbb{Z}_{3}).
\end{equation}
Instantons of both the color and horizontal gauge group can then have fractional instanton number valued in $\frac{1}{3}\mathbb{Z}$.  However due to \eqref{frac3ext} the fractional parts of the instanton numbers must be equal. More technically, the analog of equation \eqref{qcqh} relating the fractional part of the instanton numbers is now:
\begin{equation}
    \mathcal{N}_{C}=\mathcal{N}_{H},~~~\mod 1.
\end{equation}

\begin{table}[t]\centering
\large
\begin{tabular}{|c|c|c|}  \hline
 & $U(1)_{B}$  & $U(1)_{\bar{d}}$ \\ \hline
 
$SU(3)_C^{2}$ & $0$ & $N_g$ \\ \hline 

$SU(2)_L^{2}$ & $N_c N_g$ & $0$ \\ \hline 

$U(1)_Y^{2}$ & $-18 N_c N_g$ & $4 N_c N_g$ \\ \hline 

$SU(3)_H^{2}$ & $0$ & $N_c$ \\ \hline 

$[\text{CH}]$ & $0$ & $2$ \\ \hline 

\end{tabular}\caption{ABJ anomalies of chiral symmetries of quark sector with gauged $SU(3)_H$. We also show anomaly coefficients with fractional instantons, denoted as $[\text{CH}]$ (for ``color-H'' admixture), allowed by the global structure $(SU(3)_C \times 
SU(3)_H) / \mathbb{Z}_3$.}
\label{tab:Interm_anomaly}
\end{table}

To compute anomalies, we must evaluate the general sum of indices weighted by charges.  For a general flavor symmetry $f$ the anomaly coefficient$\mathcal{A}_{f}$ is:
\begin{equation}
  \mathcal{A}_{f}=\sum_{\psi_{i}} q^{f}_{\psi_{i}}I_{\psi_i}=3(\mathcal{N}_{C}+\mathcal{N}_{H})(2q^{f}_{Q}+q^{f}_{\bar{d}}+q^{f}_{\bar{u}}).
\end{equation}
From this we see that in the minimal fractional instanton, for which $\mathcal{N}_{C}=\mathcal{N}_{H}=1/3$ we have
\begin{equation}
    \mathcal{A}_{B}=0,~~~\mathcal{A}_{\bar{d}}=2.
\end{equation}
Thus fractional instantons leave the baryon symmetry, $B,$ untouched, but turn $\mathbb{Z}_{3}^{\bar{d}}$ into a non-invertible symmetry.  

As in our analysis in Section \ref{sec:masslessd} above, we now arrive at a model where the down-type yukawa coupling must vanish due to the presence of non-invertible  chiral symmetry.  Indeed, $SU(3)_{H}$ gauge invariance means that the yukawa is reduced to a single number $y_{d}$:
\beq 
\c{L}_{y_d} = y_d \delta^{i\bar{j}} H Q_i \bar{d}_{\bar{j}}.
\eeq
The down quarks transform under the non-invertible $\mathbb{Z}_{3}^{\bar{d}}$ and hence, as long as this is a good symmetry the down-type quarks are massless.

Just as in the $U(1)_{H}$ extension, we see that when the non-invertible $\mathbb{Z}_{3}^{\bar{d}}$ Peccei-Quinn symmetry is preserved there are massless quarks and CP is a symmetry of the strong sector of the SM.  Moreover again, this class of models exists only occurs for the choice of global form of the gauge group $(SU(3)_{C}\times SU(3)_{H})/\mathbb{Z}_{3}$ which in turn is only possible because $N_{c}=N_{g}.$  We now turn to a UV completion of these models which can break these symmetries and generated physical quark masses, mixing, and the CKM phase.

\section{Non-Invertible Symmetry Breaking from Color-Flavor Unification} \label{sec:fromUV}

Understanding the non-invertible chiral symmetries of the infrared theories above points us to an ultraviolet theory where small instantons dynamically break these symmetries. The minimal choice is $SU(9)$ color-flavor unification, where the $3$ colors and $3$ generations of each quark field are intermingled in the fundamental of $SU(9)$. We will describe this theory in Section \ref{sec:SU9theory} and see its $U(1)_{\rm \scriptscriptstyle PQ}$ symmetry protects down-quark masses and ensures strong CP violation vanishes in the UV. In Section \ref{sec:interBreaking} we will see that at the scale $\Lambda_9$ where $SU(9) \rightarrow \sutt$, the instantons dynamically generate a flavor-symmetric $y_d$ while keeping $\bar \theta = 0$, with further detail on the 't Hooft vertices given in Appendix \ref{app:SU9thooft}. In Section \ref{sec:flavorBreaking} we generate non-trivial flavor structure and weak CP-violation at the scale $\Lambda_3$ where $\sutt \rightarrow SU(3)_C$ while ensuring $\bar \theta$ continues to vanish, and in Section \ref{sec:scales} we discuss determining $\Lambda_9/\Lambda_3$ from the running of the strong gauge coupling. 

In the spirit of the massless up quark solution to the strong CP problem---which does not work in the SM where the instanton effects are not large enough---our ultraviolet color-flavor unified $SU(9)$ theory will contain additional instantons which can have larger effects in generating quark masses. 
Embedding the SM in such an UV structure has all the aesthetic, reductionist, conceptual appeal of grand unification, and means that the core of our mechanism has very few moving parts.
There is additional physics benefit in the possibility to separate the scale of instanton effects $\Lambda_9$ where $SU(9) \rightarrow SU(3)^2/\mathbb{Z}_3$ and the scale of flavor $\Lambda_3$ where $SU(3)_H$ is broken.
And of course this unification provides the breaking of the non-invertible chiral symmetries of quarks discussed in detail in Section~\ref{sec:fromIR}, which are the basic framework for this massless quark solution. 

However, our embedding in a flavor-unified gauge theory also makes the UV theory way too flavor-symmetric and imposes the familiar, non-trivial challenges of UV flavor model-building. 
This means that writing down a fully realistic model requires writing a predictive theory of the entire quark yukawa sector. 
This is a lofty and important goal, but for now we will attempt to factorize issues, and return to the task of flowing precisely to the SM in future work.
Of course we must ensure that we can do this breaking and generate the non-trivial yukawa matrices $y_u, y_d$ without upsetting our achievement in providing the boundary condition $\bar \theta(\Lambda_9) = 0$.
In particular this includes generating the CP-violating phase in the CKM matrix, as invariantly parametrized by Jarlskog $\tilde{J} = \text{Im}\det\left(\left[y_u^\dagger y_u, y^\dagger_d y_d\right]\right)$. 

Many approaches to strong CP protect $\bar \theta$ from $\delta_{\rm \scriptscriptstyle CKM}$ in a way that intrinsically relies on the small sizes of some entries in the CKM matrix.
In contrast, we will describe one possible, general scheme to implement flavor-breaking in which $\bar \theta$ continues to vanish without relying on the specific structure of the low-energy SM. 
We will use the gauged flavor symmetry to our advantage in recognizing that it provides a nice way to generate non-trivial complex structure in the yukawas while keeping them hermitian, $y_u^\dagger = y_u, y_d^\dagger=y_d$ (in the canonical UV basis). 
With the UV PQ symmetry producing $\bar \theta(\Lambda_9) = 0$, this method of communicating nontrivial flavor and weak CP violation to the quarks at the scale $\Lambda_3$ will guarantee that the strong CP problem continues to be solved. 
We will make a more precise statement in Section \ref{sec:flavorBreaking}.
While a fully realistic understanding of flavor in these theories will require additional work, the structure we describe shows a general scheme for solving the strong CP problem in this framework.  

\subsection{The \texorpdfstring{$SU(9)$}{SU(9)} Unified Theory and \texorpdfstring{$\bar \theta(\Lambda_{9})=0$}{theta = 0}} \label{sec:SU9theory}

An example of a unified theory which provides the $\mathbb{Z}_3$ magnetic monopoles (or $\mathbb{Z}_3$ small instantons) to break the one-form symmetry of the $SU(3)^2/\mathbb{Z}_3$ gauge theory, hence breaking the non-invertible symmetry, is the embedding in $SU(9)$ color-flavor unification. This is minimal in that it requires no new fermions, being simply a gauging of the global symmetries of the Standard Model quark fields \cite{Allanach:2021bfe}, as evinced in the fermion content given in Table \ref{tab:UVcharges}. 

\begin{table}[ht]\centering
\large
\renewcommand{\arraystretch}{1.3}
\begin{tabular}{|c|c|c|c|}  \hline
 & $SU(9)$ & $U(1)_{\tilde{B}}$ & $U(1)_{\bar d}$ \\ \hline

$\mathbf{Q}$ & $9$ & $+1$ & $0$ \\ \hline

$\mathbf{\bar u}$ & $\bar 9$ & $-1$ & $0$ \\ \hline

$\mathbf{\bar d}$ & $\bar 9$ & $0$ & $+1$ \\ \hline

$H$ & $1$ & $0$ & $0$ \\ \hline

\end{tabular}\caption{Standard Model matter content of the ultraviolet color-flavor unified gauge theory with classical global symmetries.}\label{tab:UVcharges}
\end{table}

In the UV Lagrangian we explicitly write down the `top' yukawa, and in a general basis the UV Lagrangian contains
\begin{equation}
    \mathcal{L}_0 = y_t \tilde{H} \mathbf{Q} \mathbf{\bar u} + \text{h.c.} + \frac{i \theta_9}{32\pi^2} F \tilde{F},
\end{equation}
where our notation is $\mathbf{Q} \mathbf{\bar u} = \mathbf{Q}^A \mathbf{\bar u}_A$ ($A=1,\cdots, 9$ is a $SU(9)$ index) and $F_{\mu\nu}$ is the field strength of $SU(9)$ gauge field with its dual defined as usual $\tilde{F}^{\mu\nu} = \frac{1}{2} \epsilon^{\mu\nu\rho\sigma} F_{\rho\sigma}$.
The full UV theory includes a couple more terms for reasons that we will explain in detail below. For now, we list those terms and give a brief motivation for each of them. In addition to $\mathcal{L}_0$, our UV theory includes the following additional terms.
\begin{itemize}
    \item[1.] $\mathcal{L}_\Phi = \vert D_\mu \Phi \vert^2 - V(\Phi)$

    Here, $\Phi$ is a scalar field transforming in the three-index symmetric representation of $SU(9)$. It is responsible for the breaking $SU(9) \rightarrow SU(3)^2/\mathbb{Z}_3$ at a high scale $\langle \Phi \rangle = \Lambda_9$. Even after this breaking, the gauged quark flavor symmetry $SU(3)_H$ implies that the quark yukawas continue to be flavor symmetric. 

    \item[2.] $\mathcal{L}_\Sigma = \vert D_\mu \Sigma_1 \vert^2 + \vert D_\mu \Sigma_2 \vert^2 - V (\Sigma_1, \Sigma_2 ), \;\;\; V_{\mathbb{Z}_4}(\Sigma) = \eta_1 \text{Tr} \left(\Sigma^4\right) + \eta_2 \text{Tr} \left(\Sigma^2\right)^2 + \text{ h.c.}$

    We introduce two $SU(9)$ adjoint scalars $\Sigma_1$ and $\Sigma_2$ to further break $SU(3)^2 / \mathbb{Z}_3 \rightarrow SU(3)_C$ and we assume such a breaking occurs at a scale $\langle \Sigma \rangle = \Lambda_3 \lesssim \Lambda_9$. Since reproducing the observed SM requires not only real entries of $3 \times 3$ mass matrices (quark masses and flavor mixings) but also a complex CP-violating phase ($\delta_{\scriptscriptstyle \rm CKM}$), we need to introduce CP-violating parameter(s). In our theory the vevs of $\Sigma_{1,2}$ generate the desired texture, and complex parameters in their potential $V(\Sigma)$ provide necessary CPV phases. When the theory respects $\mathbb{Z}_4$ symmetry of $\Sigma$ (which, however, is not essential for our mechanism to work and we elaborate on this below), the potential takes a simple form as shown above. There, we combined two $\Sigma_{1,2}$ to form a single ``complex'' adjoint field $\Sigma = \Sigma_1 + i \Sigma_2$ and $\eta_{1,2}$ are two complex parameters. Though note that there are additional terms but with \emph{real} parameters, e.g.~$\text{Tr} \left( \Sigma^\dagger \Sigma \right)^2$, which we have not written down because they play no key role.

    \item[3.] $\mathcal{L}_{\rho\chi} = \vert D_\mu \rho \vert^2 + i \chi^\dagger \slashed{\partial} \chi + \lambda_d \mathbf{\bar d} \rho \chi  + a_1 \rho \Sigma \rho^\dagger + a_2 \rho \Sigma \Sigma \rho^\dagger + \text{h.c.} + \rho \left( c_1 \Sigma^\dagger \Sigma + c_2 \Sigma \Sigma^\dagger \right)\rho^\dagger$ 

    Since the Jarlskog invariant $\tilde{J} \propto \text{Im} \det \left[y_u^\dagger y_u, y_d^\dagger y_d\right]$ measures the ``misalignment'' of the up vs down yukawas, generating the desired flavor structure (both real texture as well as CPV phase) requires the up- and down-type quarks are not treated identically. We implement this up-down asymmetry by introducing one completely sterile Weyl fermion $\chi$ and one $SU(9)$ fundamental scalar $\rho$ with hypercharge such that only ``down-philic'' interactions among these and the SM quarks are allowed.\footnote{We note $\chi$ has the same charge assignment as a right-handed neutrino, though it need not be one. But if it is, this model introduces no new fermions past the $16 N_g$ of a Standard Model.} $\rho$ is charged under $U(1)_{\rm \scriptscriptstyle PQ}$ but does not get a vev.
             
\end{itemize}
$SU(9)$ representations of these additional fields are summarized in Table \ref{tab:UVchargesMore}. Also, we list in Table \ref{tab:branching} how various $SU(9)$ representations decompose upon symmetry-breaking.

\begin{table}[h]\centering
\large
\renewcommand{\arraystretch}{1.3}
\begin{tabular}{|c|c|c|}  \hline
 & $SU(9)$ & $U(1)_{\tilde{B}+\bar d}$ \\ \hline

$\Phi$ & $165$ & $0$ \\ \hline

$\Sigma_{1,2}$ & $80$ & $0$ \\ \hline

$\rho$ & $9$ & $-1$ \\ \hline

$\chi$ & $1$ & $0$ \\ \hline

\end{tabular}\caption{Additional matter content used to break down to the SM in the infrared. The $SU(9)$-charged fields are all scalars, and $\chi$ is a singlet fermion. }\label{tab:UVchargesMore}
\end{table}

\begin{table}[h]\centering
\large
\renewcommand{\arraystretch}{1.3}
\begin{tabular}{|c|c|c|}  \hline
$SU(9)$ & $\sutt$ & $\suto$ \\ \hline
$9$ & $(3,3)$ & $2(3)_{+1}+(3)_{-2}$ \\ \hline
$80$ & $(8,8)+(8,1)+(1,8)$ & $5(8)_{0}+2(8)_{+3}+2(8)_{-3}+4(1)_{0}+2(1)_{+3}+2(1)_{-3}$ \\ \hline
$165$ & $(10,10)+(8,8)+(1,1)$ & \begin{tabular}{@{}c@{}} $(10)_{-6}+2(10)_{-3}+3(10)_{0}+4(10)_{+3}$ \\ $+4(8)_{0}+2(8)_{-3}+2(8)_{+3}+(1)_0$\end{tabular} \\ \hline
\end{tabular}\caption{Branching of some $SU(9)$ reps to $\sutt$ and $\suto$, where non-abelian representations are in parentheses and abelian charges are in subscripts, with the multiplicity of representations as a prefactor. }\label{tab:branching}
\end{table}

In the remaining part of this subsection, we show explicitly that strong CP violation is absent in the $SU(9)$ phase of the theory. Subsequent threshold corrections, generation of the CKM phase, and potential renormalization of $\bar \theta$ will be discussed in the following sections.  

We first note that the top yukawa explicitly breaks the separate classical global symmetries for $\mathbf{Q}, \mathbf{\bar u}$ down to the diagonal $U(1)_{\tilde{B}}$. 
This leaves two classical $U(1)$s, which at the quantum level are subject to ABJ anomalies $SU(9)^2U(1)_{\tilde{B}} = SU(9)^2U(1)_{\bar d}=1$, so
should be arranged into the familiar $SU(9)$-anomaly-free baryon number $B = \tilde{B} - \bar d$. The other direction is a flavor-unified Peccei-Quinn symmetry which we take as $\tilde{B} + \bar d$.~\footnote{We may just as well shift $\tilde{B} + \bar d$ by any good symmetry and use such a new direction as the Peccei-Quinn symmetry, for instance, $\bar d$.} Since the theory is asymptotically free, in the far UV the violation of the Peccei-Quinn symmetry by the anomaly becomes arbitrarily weak $\exp(-2\pi/\alpha) \rightarrow 0$. We assume this is a good symmetry of the UV and its only breaking is by these instanton effects.\footnote{Axion theories, which also make use of good $U(1)_{\rm \scriptscriptstyle PQ}$ symmetries, famously have a `quality problem' with quantum gravitational violations of this global symmetry posing severe fine-tuning issues. Despite also requiring a good PQ symmetry, our quality requirements are far less stringent so there is no naturalness issue here, as we discuss in Section \ref{app:quality}. }
Then, at the classical level the down yukawa $y_d$ is forbidden by the Peccei-Quinn symmetry.

In a general basis, $y_t = |y_t| e^{i \theta_t}$ is some complex number and the gauge theory has a phase $\theta = \theta_9$. We may perform a field redefinition $\mathbf{\bar u} = \mathbf{\bar u}' e^{-i \theta_t}$ to make the up yukawa real, bearing out the general EFT understanding of using spurions to count physical phases, see e.g. \cite{Grossman:2017thq}.
The $\mathbf{\bar u}$ rotation is anomalous, and so this rotation changes also the topological density term to $\theta = \theta_9 - \theta_t$. 
Since the down quark is classically massless we can then perform a rotation $\mathbf{\bar d} = \mathbf{\bar d}' e^{-i(\theta_9-\theta_t)}$ to manifestly remove the dependence of the Lagrangian on the topological density term. So there is a `canonical' basis in which the theta angle is absent and the masses are all real, i.e. there is no strong CP violation in the $SU(9)$ phase with a good $U(1)_{\rm \scriptscriptstyle PQ}$. 

Quantum mechanically, however, the Peccei-Quinn symmetry is broken by the instantons of $SU(9)$ and we will discuss this effect in Sec \ref{sec:interBreaking} along with the breaking of $SU(9)$ which dictates the small instanton scale. We give additional details in Appendix \ref{app:SU9thooft}. 

\subsection{\texorpdfstring{$SU(9)$}{SU(9)}-Breaking and Instanton Effects} \label{sec:interBreaking}

The first step of symmetry-breaking to $SU(3)^2/\mathbb{Z}_3$ may be achieved by the condensation of a three-index symmetric $\Phi^{ABC}$.
This gets a vev 
\begin{equation}
\langle \Phi^{ABC} \rangle = \Lambda_9 \varepsilon^{abc} \varepsilon^{ijk}, 
\end{equation}
where the $SU(9)$ indices are reinterpreted as multi-indices under the two $SU(3)$ factors, which are manifestly preserved $A\in \lbrace 1,2,\dots,8,9\rbrace \leftrightarrow ai\in\lbrace 11,12,\dots,23,33 \rbrace$. 
The fundamental $9$ branches to the bifundamental $3 \otimes 3$ as
\begin{equation}
    \bar u^A = \begin{pmatrix} \bar u^1 \\ \cdots \\ \bar u^9 \end{pmatrix} = \begin{pmatrix}
        \bar u^{ru} & \bar u^{gu} & \bar u^{bu} \\
        \bar u^{rc} & \bar u^{gc} & \bar u^{bc} \\
        \bar u^{rt} & \bar u^{gt} & \bar u^{bt} 
    \end{pmatrix} = \bar u^{ai},
\end{equation}
and one may usefully envision this as an `outer product' decomposition into a matrix of the quark colors and flavors. 
The different $SU(3)$ factors now act as left or right matrix multiplication, and the nontrivial global structure is seen simply because a left multiplication by an element of the center, $\omega \mathds{1}_C$ with $\omega$ a cube root of unity, commutes through and can cancel against a right multiplication by $\omega^{-1} \mathds{1}_H$. That is, for some general center rotations,
\begin{equation}
    \begin{pmatrix}
    e^{i\frac{2\pi}{3} n_h} & 0 & 0 \\ 
    0 & e^{i\frac{2\pi}{3} n_h} & 0 \\
    0 & 0 & e^{i\frac{2\pi}{3} n_h}
    \end{pmatrix}
    \begin{pmatrix}
        \bar u^{ru} & \bar u^{gu} & \bar u^{bu} \\
        \bar u^{rc} & \bar u^{gc} & \bar u^{bc} \\
        \bar u^{rt} & \bar u^{gt} & \bar u^{bt} 
    \end{pmatrix}
    \begin{pmatrix}
    e^{i\frac{2\pi}{3} n_c} & 0 & 0 \\ 
    0 & e^{i\frac{2\pi}{3} n_c} & 0 \\
    0 & 0 & e^{i\frac{2\pi}{3} n_c}
    \end{pmatrix} = e^{i \frac{2\pi}{3}(n_h + n_c)} \bar u^{ai}, 
\end{equation}
and all the matter in the theory is invariant along the $n_h = - n_c$ direction, since all the other irreps are contained in products of the fundamental and antifundamental.
The $\Phi$ branches as $165 \rightarrow (1,1) + (8,8) + (10,10)$ where the $8$ is the adjoint and the $10$ is the three-index symmetric tensor, and the entire $(8,8)$ is eaten by the gauge bosons. The SM matter fields are as in Table \ref{tab:inter_charges}.

Since the $SU(9)$ theory is asymptotically free, the dominant contribution to the 't Hooft vertices arises at the scale $\Lambda_9$ where the breaking $SU(9)\rightarrow SU(3)^2/\mathbb{Z}_3$ occurs. 
Note also that across $\Lambda_9$ the gauge couplings are non-trivially matched as $1/g^2_3 = 3/g_9^2$, as there is a non-trivial `index of embedding' \cite{Csaki:1998vv,Intriligator:1995id,Cordova:2023her} of $SU(3)^2/\mathbb{Z}_3$ into $SU(9)$. 
Non-trivial index of embedding means the following. Given a breaking of a gauge group $G \to H$, if the index of embedding is greater than 1, then not all of $G$-instanton effects are captured by unbroken $H$-instantons. 
For us, the index of embedding is 3. This means that there are $SU(9)$ instantons which appear to be $\mathbb{Z}_3$ fractional instantons of $SU(3)^2/\mathbb{Z}_3$ theory, as our analysis in Section \ref{sec:fromIR} showed could occur. 
These are precisely what we need to explicitly break the non-invertible symmetries.
The non-trivial index of embedding not only increases the size of UV instanton effects, but also decreases the number of legs of the 't Hooft vertex, and together these effects are crucial for our solution.

\begin{figure}
  \centering
  \includegraphics[width=.5\textwidth,trim={0 0.5cm 0 0},clip]{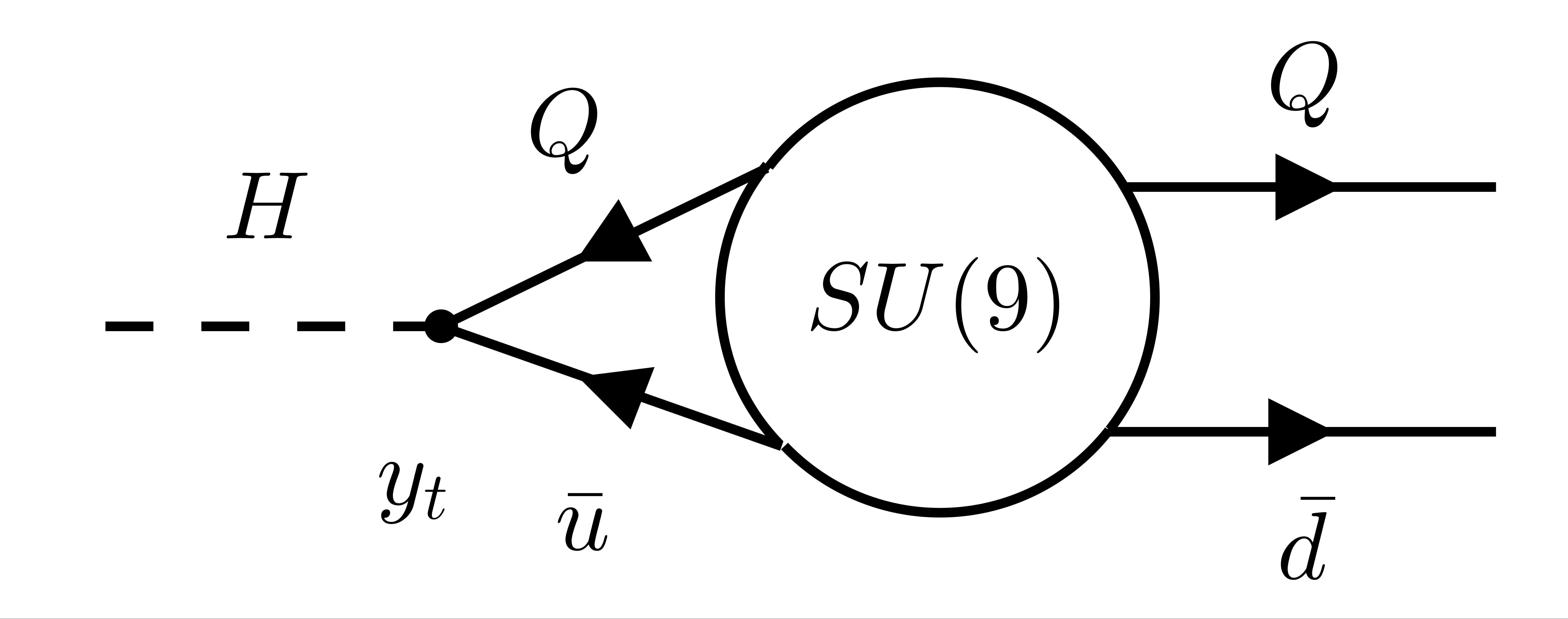}\hspace{.5cm}
  \caption{'t Hooft vertex in the $SU(9)$ theory by which instantons generate the down-type yukawa from the up-type yukawa. }
  \label{fig:SU9Inst}
\end{figure}

The instantons of the $SU(9)$ theory, in violating the anomalous global symmetries, generate the 't Hooft vertex of Figure \ref{fig:SU9Inst}.
Since we avoided adding any additional $SU(9)$-charged fermions, the 't Hooft vertices include only the SM fermion zero modes. However, the additional charged scalars will affect the instanton density and we defer a fuller discussion of the sizes of the effects of 't Hooft vertices to Appendix \ref{app:SU9thooft}. Here we content ourselves with the rough result that instantons generate a down-type yukawa which violates $U(1)_{\tilde{B}+\bar d}$ PQ symmetry by two units. Flowing down in energies and momentarily staying at some scale $\Lambda > \Lambda_9$, instantons begin to generate 
\begin{equation}
    \mathcal{L}(\Lambda) \sim y_t H \mathbf{Q} \mathbf{\bar u} +  y_t^\star e^{i \theta_9} e^{-\frac{2\pi}{\alpha_9(\Lambda)}} H \mathbf{Q} \mathbf{\bar d}  + \text{ h.c.} + \frac{i \theta_9}{32\pi^2} F \tilde{F},
\end{equation}
We see that a color-flavor symmetric down-type yukawa is generated with rough size
\begin{equation}
    y_d \sim y_t^\star e^{i \theta_9} e^{-\frac{2\pi}{\alpha_9(\Lambda)}}.
\end{equation}
A key point is that such an instanton-induced yukawa comes with just the right phase to ensure 
\begin{equation}
    \bar \theta = \arg e^{- i \theta_9} \det  y_u y_d = -\theta_9 + \arg |y_t|^2 e^{i\theta_9} = 0,
\end{equation}
where here because of the color-flavor gauge symmetry the yukawas are simply numbers. 
This seeming conspiracy among phases is guaranteed by the good PQ symmetry of the UV. In the canonical basis the statement becomes that the yukawas remain real.

Now going further down to the theory at the matching scale, in a general basis we have
\begin{equation} \label{eqn:genBottomYuk}
    \mathcal{L}(\Lambda_9) \sim y_t H \mathbf{Q} \mathbf{\bar u} +  y_t^\star e^{i \theta_9} e^{-\frac{2\pi}{3 \alpha_s(\Lambda_9)}} H \mathbf{Q} \mathbf{\bar d}  + \text{ h.c.} + \frac{i 3 \theta_9}{32\pi^2} \left( G \tilde{G} + K \tilde{K}\right) ,
\end{equation}
where $\alpha_s(\Lambda_9) = \alpha_9(\Lambda_9)/3$ is the QCD coupling evolved from the infrared up to the $\Lambda_9$ scale, and here $G, K$ are the $SU(3)_C, SU(3)_H$ gauge field strengths respectively. 
The nontrivial matching of the theta angles accounts for the yukawas being upgraded from single numbers to $3 \times 3$ matrices (from the perspective of $SU(3)_C$) and ensures 
\begin{equation}
    \bar \theta = - 3 \theta_9 + \arg \det |y_t|^2 e^{i \theta_9} = 0,
\end{equation}
which again in the canonical basis reduces to the statement that the yukawas are all real, and so have real eigenvalues.
This is the core of the massless quark solution to the strong CP problem \cite{Georgi:128245,Kaplan:1986ru,Choi:1988sy,Srednicki:2005wc}.

In contrast to our generation of Dirac neutrino masses from the charged lepton masses \cite{Cordova:2022fhg}, here the required suppression from top to bottom yukawa is not so large, $y_b/y_t \sim 1/40$. From the na\"{i}ve one-instanton \eqref{eqn:genBottomYuk} one estimates $\alpha_s(\Lambda_9) \simeq 0.57$, and we comment on the effects of quadratic fluctuations around this solution in Appendix \ref{app:SU9thooft}.
In the end, it is difficult to get a reliable analytic estimate of the instanton effects for our $SU(9)$ theory. Just as with the original massless up quark solution, lattice simulations will be needed to determine precisely how well this works. But we are aided by the gauge coupling jumping up due to the non-minimal index of embedding at the $SU(9)$ scale, and we also have a natural model-building handle to slow down or reverse its one-loop running through colored particles with masses below $\Lambda_9$, to be discussed further in Section \ref{sec:scales}.

\subsection{Flavor-Breaking and Keeping \texorpdfstring{$\bar \theta \approx 0$}{theta = 0}} \label{sec:flavorBreaking}

Having described a color-flavor unified theory which guarantees $\bar \theta = 0$ in the ultraviolet, we need to understand how $\delta_{\rm \scriptscriptstyle CKM} \sim \mathcal{O}(1)$ may appear without spoiling it. 
As discussed above, making use of gauged flavor symmetry forces us to confront the generation of the non-flavor-symmetric SM yukawa sector. 
Here we pursue the simplest possibility of a single further symmetry-breaking step $(SU(3)_C \times SU(3)_H)/ \mathbb{Z}_3 \rightarrow SU(3)_C$, in which we break the horizontal symmetry all at once at a lower scale $\Lambda_3 < \Lambda_9$. Our strategy to not destabilize our UV achievement will be to communicate flavor- and CP-breaking to the SM quarks in a way that keeps the yukawas hermitian. Hermiticity of the quark yukawas has been used in different ways for the strong CP problem before, for example in parity-symmetric theories \cite{Mohapatra:1978fy} or with supersymmetry \cite{Hiller:2001qg,Hiller:2002um} or in an effective 2HDM \cite{Evans:2011wj}, but our usage will be quite novel. In a theory of gauged flavor this will be a natural possibility as we will see below.

This separation of the scales at which the instantons generate non-invertible symmetry-violation and at which the flavor structure is generated is a possibility in this theory which differs from the theories considered in \cite{Agrawal:2017evu,Agrawal:2017ksf}.~\footnote{Of course it is true that the generation of $\Lambda_3 \ll \Lambda_9$ would constitute a hierarchy problem. But note that, unlike for the electroweak hierarchy problem, the classical natural solutions applied to this hierarchy problem still work fine. Perhaps the real role of supersymmetry or warped extra dimensions in our universe is to stabilize the scale at which flavor is generated $\Lambda_3$.} 
Generally this could aid in keeping $\bar \theta$ small, but here we will describe a mechanism to generate flavor which automatically preserves $\bar \theta = 0$ without utilizing this structural possibility. 
As seen in the prior sections, we have begun with yukawas proportional to the identity matrix $y_u, y_d \propto \mathds{1}$. Generating the non-trivial flavor structure, especially flavor hierarchies, using higher-dimensional operators suppressed by powers of $\Lambda_3 / \Lambda_9$ means that there is a tension between reproducing the SM flavor structure and having a large ratio of scales in this scheme.\footnote{Vector-like quarks would be the obvious way to address this issue, but these can be disastrous for our strong CP solution---they easily provide new sources of CP-violation which is directly communicated to the SM quarks, and in any case make our 't Hooft vertices more complicated and potentially lead to further suppression. So we forego them and utilize a strategy of communicating flavor-breaking purely bosonically.} That is, we will not ask for any large hierarchy between these two scales, nor will we in this work discuss a predictive theory of their origins. 

There may be many choices of how to break $SU(3)_H$ and match onto the SM. 
We choose as our symmetry-breaking sector two $SU(3)_H$ adjoint scalar fields $\Sigma_{1,2}$ which can together entirely break $SU(3)_H \rightarrow \varnothing$. We will find it useful to join these together into the `complex adjoint' $\Sigma \equiv \Sigma_1 + i \Sigma_2$, which can be seen merely as an accounting measure to keep track of their would-be $SO(2)\simeq U(1)_\Sigma$ global symmetry. Their nonzero commutator $\left[\Sigma_1,\Sigma_2\right] = \left[\Sigma^\dagger,\Sigma\right]/(2i)$ is required to fully break the $SU(3)_H$ symmetry. And as the SM Jarlskog invariant is written in terms of a commutator of flavor spurions it will be proportional to this single nonvanishing commutator in this model.

With only the SM fermions, there are no renormalizable interactions allowed with $\Sigma$, and its breaking of $SU(3)_H$ is communicated to the quarks solely through the broken $SU(9)$ gauge bosons.\footnote{To match onto the perturbative QCD coupling having started with large $SU(9)$ instanton effects needed to generate $y_b/y_t$, we assume all the $SU(9)$ partners of $\Sigma$ and $\Phi$ get masses at the scale $\Lambda_3$, rather than $\Lambda_9$. This way we can benefit from the rapid RG running of gauge coupling from the strong to weak regime as the theory flows from the UV $SU(9)$ phase to $SU(3)_C$ in the IR. We discuss this point in detail in Section~\ref{sec:scales}.} As these flavor-breaking effects must include the generation of the CP-violating $\delta_{\rm \scriptscriptstyle CKM}$, we must ensure that this scalar sector can break CP. Indeed, the most general potential for $\Sigma$ includes many CP-violating phases, and for simplicity we can get a more-tractable potential by imposing a $\mathbb{Z}_4$ symmetry,
\begin{equation}
    V_{\mathbb{Z}_4} (\Sigma) = \eta_1 \text{Tr} \left(\Sigma^4\right) + \eta_2 \text{Tr} \left(\Sigma^2\right)^2 + \text{ h.c.} + \text{terms with real coeffs},
\end{equation}
where we have left off the terms which do not break $U(1)_\Sigma$.
This potential has a single CP-odd phase, which is captured by the field-redefinition-invariant $\eta_1^\dagger \eta_2$. We assume this has some random complex phase, explicitly breaking CP.
In this simplified scenario there is no spontaneous violation of CP when $\Sigma$ gets a vev because $\eta_1, \eta_2$ both have charge $-4$ under the spurious $U(1)_\Sigma$ symmetry \cite{Haber:2012np}.
In general without imposing the $\mathbb{Z}_4$, the $\Sigma$ potential will both explicitly and spontaneously violate CP. 
Either way is fine; the mechanism we will now describe works however CP violation appears in $\Sigma$'s potential and we will explore the spontaneous case further in forthcoming work \cite{Koren:2024c}.

\begin{figure}
    \centering
    \includegraphics[width=0.5\linewidth]{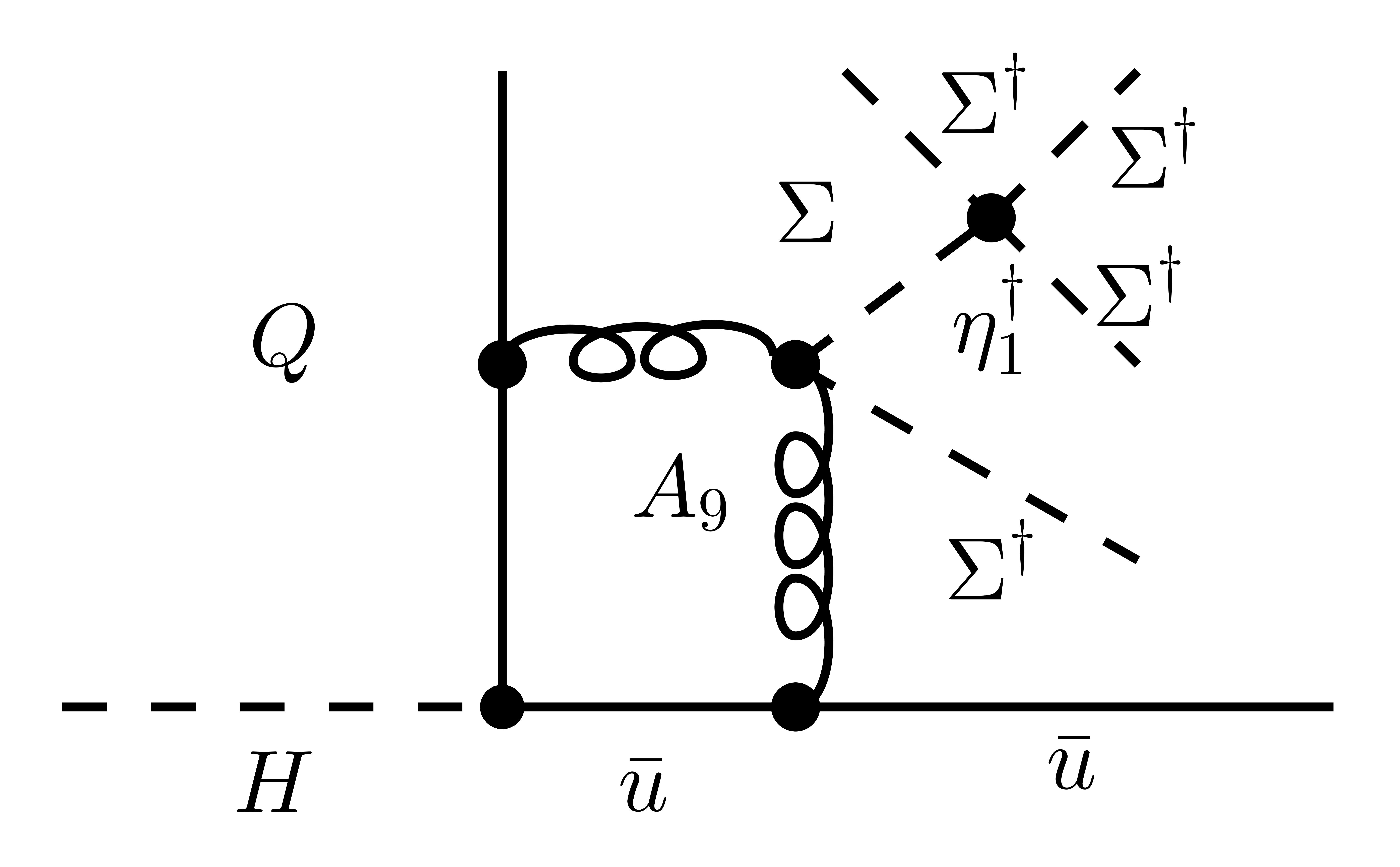}
    \caption{A $\Lambda_9$ threshold contribution to the up-type yukawas which eventually gives flavor- and CP-breaking proportional to $\alpha_9 \eta_1^\dagger \langle \Sigma^\dagger\rangle^{4a}_{\ \ b}/(4\pi)$. The hermitian conjugate contribution is generated by the same diagram with an $\eta_1$ insertion on the conjugate scalar leg, so they sum to yield a hermitian up-type yukawa. Here $\Sigma$ is the $SU(3)_H$ adjoint.}
    \label{fig:SU9CPCorrection}
\end{figure}

Despite the fact that we now have flavor- and CP-breaking, in this model $\bar \theta = 0$ continues to hold. This is because the effects of $\Sigma$ (both in producing the yukawa texture and in CP-violation) are transferred to the SM fields only through the $SU(9)$ heavy gauge bosons.
Resultingly, for any coupling one could attach to an external $\Sigma$ leg to get a complex contribution to a yukawa coupling, there is always also the hermitian conjugate coupling to be attached to the $\Sigma^\dagger$ leg. See for example Figure \ref{fig:SU9CPCorrection}. 
As a whole, our mechanism works in a way that corrections to the yukawas always leave them hermitian. Note that this argument applied just as well to the RG evolution during the $\sutt$ phase when the $\Sigma$ fields appear only in closed loops---while there are many-loop diagrams proportional to $\eta_1^\dagger \eta_2$, there are compensating diagrams proportional to $\eta_2^\dagger \eta_1$ which sum up to real yukawas in the $\sutt$ phase when they are still just numbers.

Then below the $SU(3)$-breaking scale $\Lambda_3$, the basic structure of the up-type yukawas, for example, is given by
\begin{align}
    (y_u)^a_{\ b} &\sim y_t \left(\mathds{1}^a_{\ b} + \frac{\alpha_9}{(4\pi)} \frac{\lbrace \Sigma^\dagger, \Sigma \rbrace^a_{\ b}}{2 \Lambda_9^2} + \text{other terms with real coefficients}\right. \\
    &\left. + \frac{\alpha_9} {(4\pi)}\frac{\eta_1^\dagger (\Sigma^{\dagger 4})^a_{\ b} + \eta_2^\dagger \text{Tr}(\Sigma^{\dagger 2}) (\Sigma^{\dagger 2})^a_{\ b}}{\Lambda_9^4} + \frac{\alpha_9} {(4\pi)}\frac{\eta_1 (\Sigma^{4})^a_{\ b} + \eta_2 \text{Tr}(\Sigma^{ 2}) (\Sigma^{ 2})^a_{\ b}}{\Lambda_9^4} + \dots \right), \nonumber
\end{align}
where we have only given some na\"{i}ve power counting for a notion of the size of these effects, and it should be understood that these are vevs of $\Sigma$. 
The anticommutator $\lbrace \Sigma^\dagger, \Sigma \rbrace/2 = \Sigma_1^2 + \Sigma_2^2$ is the structure that appears from the gauge interactions, and these must respect the $U(1)_{\Sigma}$ spurious `flavor' symmetry. Then this is the leading correction to the flavor structure which is manifestly real, all of which we group on the top line. 
In the second line we give the first corrections through which a complex phase enters the yukawas in the simplified scenario where the $\mathbb{Z}_4$ symmetry of $\Sigma$ controls the potential. 

So far the corrections consist of a sum over diagrams attaching a vertex with complex coefficient to the $\Sigma$ line and from attaching its conjugate vertex to the $\Sigma^\dagger$ line. 
Then while the generated yukawas are no longer real, they remain hermitian, which suffices to guarantee $\det y_u, \det y_d \in \mathbb{R}$, and so $\bar \theta = 0$.  
Note also that while in the unbroken phase the CP violation was necessarily invariantly parametrized only by $\eta_1^\dagger \eta_2$, at low energies after $\Sigma$ gets a vev we integrate out its fluctuations and it appears only as an external source, so also the combination  $\eta_1^\dagger \text{Tr} \langle \Sigma \rangle^{\dagger 4} + \eta_2^\dagger \text{Tr}(\langle \Sigma \rangle^{\dagger 2})^2$ can invariantly parametrize CP-violation. 

However, while we now have complex, $SU(3)_H$-violating yukawas, this setup cannot yet generate the CKM CP-angle. The issue is that the $SU(9)$ dynamics affect the up and down quarks symmetrically, whereas a nonvanishing CKM angle appears from a mismatch in the form of the yukawa couplings, $\sin \delta_{\rm \scriptscriptstyle CKM} \propto \text{Im}\det\left(\left[y_u^\dagger y_u, y^\dagger_dy_d\right]\right)$. So we must introduce another ingredient to skew the yukawas apart from each other, while not upsetting the solution. 

\begin{figure}
    \centering
    {{\includegraphics[width=0.45\linewidth]{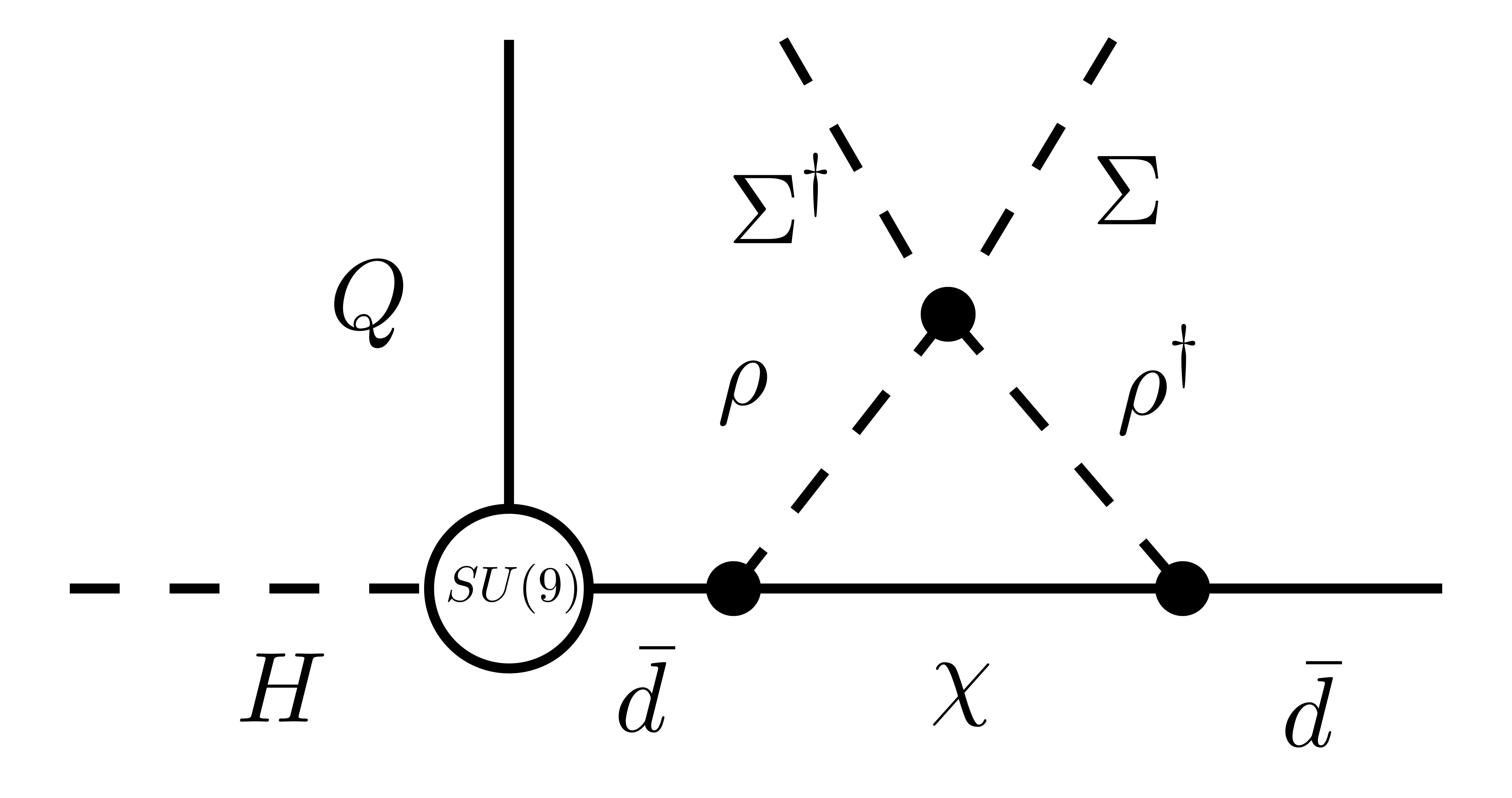}}}%
    \qquad
    {{\includegraphics[width=0.45\linewidth]{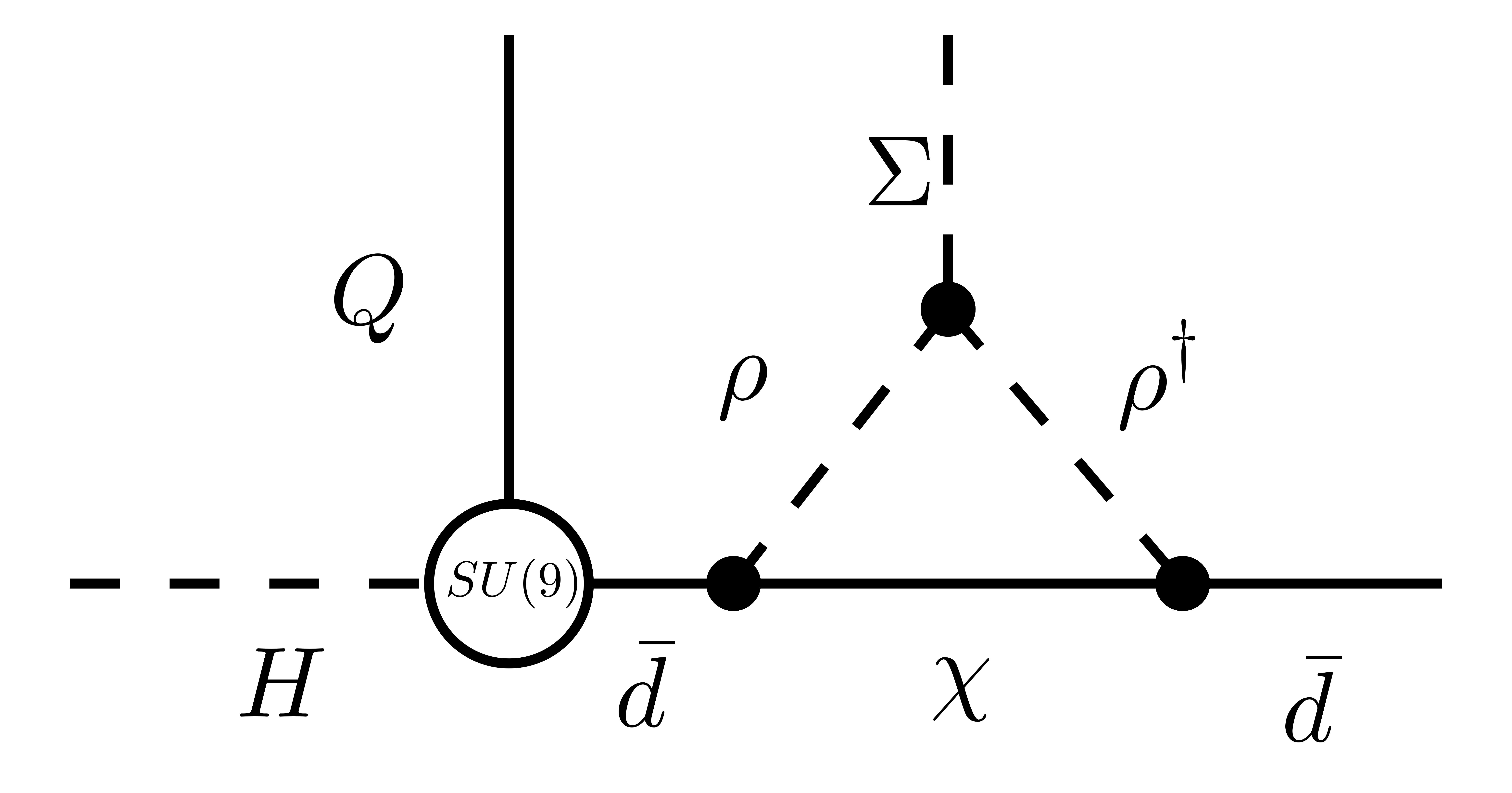}}}%
    \caption{Threshold corrections to the down-type yukawas. The left diagram preserves the $\mathbb{Z}_4$, and gives a flavor-breaking but CP-preserving contribution proportional to $|\lambda_d|^2 c_1 \langle \Sigma^\dagger \Sigma \rangle^{a}_{\ \ b}$. The right diagram is present in the general $\mathbb{Z}_4$-violating case, and gives flavor- and CP-breaking proportional to $|\lambda_d|^2 a_1 \langle \Sigma \rangle^{a}_{\ \ b}$. Again, the hermitian conjugate contribution is generated by the same diagram with an $a_1^\dagger$ insertion of $\Sigma^\dagger$, so they sum to yield a hermitian down-type yukawa. These $\rho$-mediated corrections appear solely for the down-type quarks.}
    \label{fig:SquarkCPCorrection}
\end{figure}

In particular we can introduce some fields which interact only with, say, the down quark and not the up quark. This is simplest if it does not allow for any new CP phases in operators containing quarks nor introduce any new color-flavored fermions which would appear in our 't Hooft vertices. With two new fields $\rho$ and $\chi$ a new yukawa can be allowed, where one allocation of quantum numbers is for the scalar $\rho$ to be a down squark and the fermion $\chi$ to be sterile. The scalar can furthermore couple directly to $\Sigma$,
\begin{align}
        \mathcal{L}_{\rho\chi} \supset & \ \lambda_d \mathbf{\bar d} \rho \chi  + a_1 \rho \Sigma \rho^\dagger + a_2 \rho \Sigma \Sigma \rho^\dagger + h.c.  \\ &+ \rho \left( c_1 \Sigma^\dagger \Sigma + c_2 \Sigma \Sigma^\dagger \right)\rho^\dagger, \nonumber
\end{align}
where we have suppressed indices to avoid notational clutter, e.g. $\rho \Sigma \Sigma \rho^\dagger = \rho_a \Sigma^a_b \Sigma^b_c (\rho^\dagger)^c$.
These `down-philic' interactions generate a loop correction to the down type yukawas which is not present for the up type yukawas. As we now discuss, this allows a CKM phase to be generated. We can use a $\chi$ rotation to make $\lambda_d$ real, and $c_{1,2}$ are real by self-hermiticity of the operators. $a_{1,2}$ are set to zero if the $\mathbb{Z}_4$ is imposed, or in general have complex phases and lead to further field-redefinition-invariant CP-odd parameters such as $a_1^2 a_2^\dagger$ or $\eta_1^\dagger a_2^2$.
In either case, the interactions of $\rho$ to $\Sigma$ allow further flavor-violation to be communicated to the quarks in a way that does not upset our mechanism: $\Sigma$ couples to $\rho^\dagger \rho$, such that it will always enter in a hermitian manner. See Figure \ref{fig:SquarkCPCorrection}.

Overall, this mechanism works by having all CP-violating phases in the scalar sector and communicating them to the SM quark sector via a bosonic mediation. 
In particular the interactions of the $\Sigma$ fields with the mediators (who will transfer the flavor-and CP-breaking to the SM quarks) are always hermitian in the mediator fields, though not necessarily in the $\Sigma$ fields.
Then the requirement that the Lagrangian is hermitian itself ensures that for any diagram with possibly complex phases in it, there always exists a conjugate diagram where the $U(1)_\Sigma$ charges are all reversed and you get the complex conjugate phase. 
This would not be true if the CP-violating phases were directly coupled to the SM fermions since then conjugating the diagram requires charge conjugating the fermion legs, resulting in a different diagram from the original one. 
Instead here we automatically get a sum over all $\Sigma$ and $\Sigma^\dagger$ source insertions when computing the yukawa corrections, which keeps them hermitian.

Now let us examine the form of the yukawas and the CP-violating phases we have generated. 
Recall that, aside from the strong CP angle itself, the only field redefinition invariant CP-odd parameter in the theory of quarks is the Jarlskog invariant,
\begin{equation}
    \tilde{J} = \text{Im} \det \left( \left[ y_u^\dagger y_u, y_d^\dagger y_d\right]\right).
\end{equation}
This parameter is measured to be small, $\tilde{J} \simeq \frac{m_c}{m_t}\frac{m_s}{m_b} J \sim 5 \times 10^{-9}$.

Since the only object breaking the flavor symmetry is $\Sigma$, the CKM phase must depend on $\left[ \Sigma, \Sigma^\dagger \right]$ as the only nonvanishing commutator. Now working even more schematically just to see the structure, in the $\mathbb{Z}_4$ conserving case we have (see the left panel of Figure \ref{fig:SquarkCPCorrection})
\begin{align}
    y_u &\sim \mathds{1} + \lbrace \Sigma^\dagger, \Sigma \rbrace + \left(\eta \Sigma^4 + \eta^\dagger \Sigma^{\dagger 4}\right) + \dots \\
    y_d &\sim \mathds{1} + c \Sigma^\dagger \Sigma + \lbrace \Sigma^\dagger, \Sigma \rbrace + \left(\eta \Sigma^4 + \eta^\dagger \Sigma^{\dagger 4}\right) \dots,
\end{align}
where we have only written down enough terms to see both that CP-violation enters (through the $\eta$ complex self-couplings of $\Sigma$) and that the two structures differ (due to the $c$ couplings of $\rho$ to $\Sigma$). Moreover, we have absorbed the $\Lambda_9^{-1}$ into the scalar field, and we have left off most of the constants including the overall proportionality factors as $y_u \propto y_t$ and $y_d \propto y_b$ where $y_b \sim y_t^\star e^{i\theta_9} e^{- 2\pi / \alpha_9 (\Lambda_9)}$. As discussed above we can choose a basis where both are real numbers and for convenience we do so. To analyse the Jarlskog invariant it can be useful to split $y_d = r y_u + r \Delta y$ into a piece which is simply a real rescaling ($r \in \mathbb{R}$) of the $y_u$ structure (so will manifestly commute) and an extra piece which for us is given by the extra effects of the $\rho, \chi$ interactions. For us, $r \sim e^{- 2\pi / \alpha_9 (\Lambda_9)}$ and $\Delta y \sim c \Sigma^\dagger \Sigma$ with $c \in \mathbb{R}$. Then we have
\begin{align}
    \tilde{J} &= \text{Im} \det \left( r^2 \left[ y_u^2,  \lbrace y_u, \Delta y\rbrace + \Delta y^2 \right]\right) \\
    &\supsetsim \text{Im} \det \left( 4 r^2 \left[ \eta \Sigma^4 + \eta^\dagger \Sigma^{\dagger 4}, c \Sigma^\dagger \Sigma \right] \right) + \dots,
\end{align}
where in the second line we only show the lowest-order structure that can contribute to weak CP violation assuming $\mathbb{Z}_4$ invariance. In general without the $\mathbb{Z}_4$ this could come in at a lower order. The $r^2$ dependence reflects the fact that $\tilde{J}$ is proportional to $y_d^2$, while $c$-dependence shows that $\tilde{J}=0$ without up-down asymmetry factor. 
We can rewrite this in terms of the commutator 
$\left[\Sigma,\Sigma^\dagger\right]$ as
\begin{align}
    \tilde{J} \propto \text{Im} \det \bigg( &\eta \left( \left[\Sigma,\Sigma^\dagger\right] \Sigma^4 + \Sigma \left[\Sigma,\Sigma^\dagger\right] \Sigma^3 + \Sigma^2 \left[\Sigma,\Sigma^\dagger\right] \Sigma^2 + \Sigma^3 \left[\Sigma,\Sigma^\dagger\right] \Sigma \right) \\
     - & \eta^\dagger \left( \Sigma^{\dagger} \left[\Sigma,\Sigma^\dagger\right] \Sigma^{\dagger 3} + \Sigma^{\dagger 2} \left[\Sigma,\Sigma^\dagger\right] \Sigma^{\dagger 2} + \Sigma^{\dagger 3} 
 \left[\Sigma,\Sigma^\dagger\right] \Sigma^{\dagger} + \Sigma^{\dagger 4} \left[\Sigma,\Sigma^\dagger\right] \right) \bigg). \nonumber \\
 = \text{Im} \det \bigg( &\eta \left( \left[\Sigma,\Sigma^\dagger\right] \Sigma^4 + \Sigma \left[\Sigma,\Sigma^\dagger\right] \Sigma^3 + \Sigma^2 \left[\Sigma,\Sigma^\dagger\right] \Sigma^2 + \Sigma^3 \left[\Sigma,\Sigma^\dagger\right] \Sigma \right) - \text{h.c.} \bigg) \nonumber
\end{align}
This will be generically nonvanishing so long as $\eta \notin \mathbb{R}$, and $\left[\Sigma,\Sigma^\dagger\right] \neq 0$ which is anyway necessary for $\langle \Sigma \rangle$ to properly break the horizontal symmetry.

Therefore, we have succeeded in generating the weak CP phase of the quarks, while keeping the strong phase vanishing due to the hermiticity. In a more-general case that does not preserve the $\mathbb{Z}_4$, the power counting can come out differently. For example because we can rely on CP violation in the coupling $\rho^\dagger ( a \Sigma + a^\dagger \Sigma^\dagger) \rho$ rather than in only $\Sigma$ self-couplings. This will yield
\begin{align}
    \tilde{J} \propto \text{Im} \det \bigg( &a^\dagger \left( \left[\Sigma,\Sigma^\dagger\right] \Sigma^\dagger + \Sigma^\dagger \left[\Sigma,\Sigma^\dagger\right] \right) 
    - a \left( \Sigma \left[\Sigma,\Sigma^\dagger\right] + \left[\Sigma,\Sigma^\dagger\right] \Sigma \right) \bigg),
\end{align}
and while still suppressed by the gauge coupling, the bottom yukawa, and two loop factors as before, we have less suppression now by a factor of $\sim (\Lambda_3/\Lambda_9)^3$.

Let us review the story to discuss what value of $\bar \theta$ these models predict. In the $SU(9)$ phase above $\Lambda_9$, there was no strong CP violation as a result of the $U(1)_{\rm \scriptscriptstyle PQ}$. At $\Lambda_9$, instantons generated PQ-violation but ensured $\bar \theta = 0$, and in the $\sutt$ phase the yukawas are simply numbers so there is no other CP-odd quark parameter. The RG evolution in this phase and the matching at $\Lambda_3$ when $\Sigma$ gets a vev both have a structure that generates complex, but hermitian yukawas, keeping $\bar \theta = 0$. Only below $\Lambda_3$, after integrating out $\Sigma$, is the CKM angle $\delta_{\rm \scriptscriptstyle CKM}$ present to renormalize $\bar \theta$. But now we are back to the SM field contents, and so this renormalization is miniscule and protected by the SM structure, with the finite renormalization producing only $\bar \theta \sim 10^{-16}$ as estimated by Ellis \& Gaillard \cite{Ellis:1978hq}.

Before concluding this section on the $SU(3)_H$-breaking, let us emphasize again what we have achieved. Often, ensuring that models designed to solve the strong CP problem do not end up generating unacceptably large $\bar \theta$ relies explicitly on the small entries of the SM yukawa matrices. Recall this is how it works in the SM itself, as many loops are needed to generate $\bar \theta$ from $\delta_{\rm \scriptscriptstyle CKM}$, so it is sensible to model-build toward that same conclusion. 
One such example is the flavor deconstructed massless quark solution of \cite{Agrawal:2017evu}.
However, in this work we started with the flavor symmetry fully gauged and we need to generate the non-trivial yukawas. Then utilizing small quark mixing angles to keep $\bar \theta$ small is more challenging for us since it would require us first to construct a predictive theory of flavor, i.e. specify the full texture of $\langle \Sigma \rangle$ and scalar potential compatible with observations. 

Instead, we propose a mechanism which factorizes the issue of strong CP from the precise details of the SM flavor structure. Namely, our tactic has been to describe a way to generate non-trivial flavor structure and weak CP violation which automatically does not produce nonvanishing $\bar \theta$. 
When the flavor structure of $\langle\Sigma\rangle, \langle\Sigma^\dagger\rangle$ is communicated to the SM quarks in the way described above, the yukawas continue to be hermitian and $\bar \theta = 0$ holds, no matter what the flavor structure is.
That is, this solution would continue to work even if the quark sector were fully anarchic and the yukawas were described only with $\mathcal{O}(1)$ numbers.
Furthermore, it has not escaped our notice that this mechanism has an immediate application to a flavorful theory of spontaneous CP violation which we will report on in forthcoming work \cite{Koren:2024c}.

\subsection{Scales and Running} \label{sec:scales}

We would like to get an idea of the scale $\Lambda_9$ at which color-unification occurs, which is dictated by the gauge coupling $\alpha_9(\Lambda_9)$ being the right size to generate the bottom yukawa from the top yukawa. Since we need relatively large gauge coupling, this is subject to theoretical uncertainty and lattice simulations are needed to accurately determine the size of instanton effects. Furthermore, the evolution of the gauge coupling is dictated by the entire charged matter spectrum, which depends both on the representations we have added to implement flavor-breaking and also on their masses, so in principle there is a lot of freedom.

Across the breaking scale $\Lambda_9$, $SU(3)_C$ is embedded non-trivially into $SU(9)$ with an index of embedding of $3$, such that the gauge coupling is rescaled as 
\begin{equation}
    \alpha_9(\Lambda_9) = 3 \alpha_s(\Lambda_9).
\end{equation}
This relative factor of three strengthens the small $SU(9)$ instanton effects with respect to those at lower energies, which is one reason they may achieve some qualitatively new effects. But as discussed above the needed gauge coupling is large and there may need to be additional running between the $SU(9)$ breaking scale $\Lambda_9$ and the scale $\Lambda_3$ at which we break to $SU(3)_C$. 

\begin{figure}
    \centering
    \includegraphics[width=0.6\linewidth]{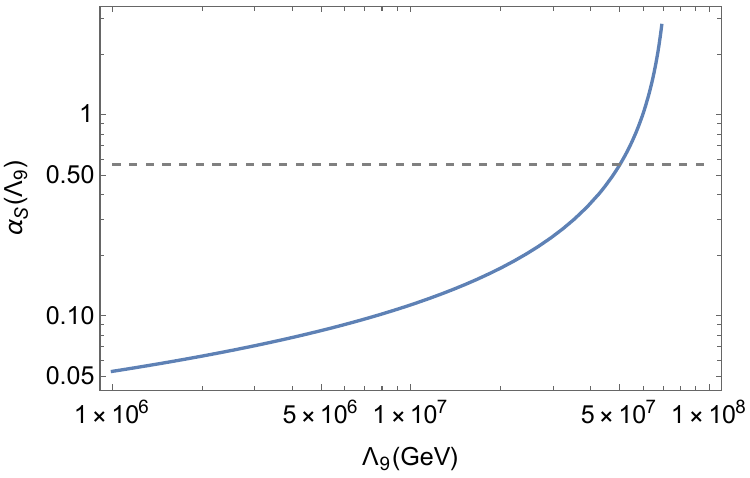}
    \caption{The strong coupling at the matching scale in a simplified scenario where the colored scalars all have a common mass $\Lambda_3 = 10^6 \text{ GeV}$. At $\Lambda_9$ the gauge coupling jumps by the index of embedding $\alpha_9(\Lambda_9)=3\alpha_S(\Lambda_9)$ but above $\Lambda_9$ the theory is again asymptotically free. For some very rough guidance the dashed line is the na\"{i}ve estimate of \eqref{eqn:genBottomYuk} to generate the bottom yukawa.}
    \label{fig:alpha9running}
\end{figure}

Above $\Lambda_9$ the $SU(9)$ theory should be  asymptotically free, which is easily achieved because of the large number of colors. 
At the scale $\Lambda_3$ we match onto the SM, and heavy gauge bosons generate flavor-changing four-fermi operators which are tightly constrained, so we likely must have $\Lambda_3 \gtrsim 1000 \text{ TeV}$ \cite{Isidori:2010kg,Ellis:2020unq}. 
In between these scales, the theory must switch to being IR-free such that the coupling grows into the UV by the beta function 
\begin{align}
    \alpha_s^{-1}(\Lambda_9) &\simeq \alpha_s^{-1}(\Lambda_3) + \frac{\beta_{3}}{2\pi} \log\left(\frac{\Lambda_9}{\Lambda_3}\right) \\
    \beta_3 &= \left(\frac{11}{3} N_c - \frac{4}{3} n_f I_f - \frac{1}{3} n_s I_s - \frac{1}{6} n_r I_r \right),
\end{align}
where the charged matter consists of $n_f$ Dirac fermions, $n_s$ complex scalars, and $n_r$ real scalars with Dynkin indices $I_f, I_s, I_r$ respectively.
The SM has $n_f = 2 N_g$ fundamental Dirac fermions. We normalize the $SU(N)$ generators so that the Dynkin index for the fundamental representation is $1/2$, giving $I_f = 1/2$. 
Our breaking sector as outlined in Table \ref{tab:UVchargesMore} has added $2 \times (1+8)$ adjoint scalars with $I_r = N_c$ coming from $\Sigma$, $10$ three-index symmetric scalars with $I_s = 15/2$ coming from the components of $\Phi$ which were not eaten, and $N_g$ fundamental scalars with $I_s = 1/2$ from $\rho$. This is far more than enough colored matter to overpower the gluonic contribution if these scalars for some reason do not become massive until a scale below $\Lambda_9$. Then one only needs an extremely mild hierarchy for appreciable running to take place, with $\beta_3 = -55/2$.

From the bottom-up we have precisely measured the low-energy value of the strong coupling, and the PDG gives the world-average $\alpha_s(M_Z) \approx 0.118$ \cite{ParticleDataGroup:2022pth}. From the scales $M_Z$ up to $\Lambda_3$ we have the SM degrees of freedom, which contribute to a beta function $\beta_{\scriptscriptstyle \rm QCD} = (11 - 2/3 n_f)$ with $n_f=5$ below $m_t \approx 173 \text{ GeV}$ and $n_f = 6$ above.

For simplicity, we consider a scenario where all of the new colored scalar degrees of freedom get masses only at $\Lambda_3 = 10^6 \text{ GeV}$, and we plot in Figure \ref{fig:alpha9running} the relationship between $\Lambda_9$ and $\alpha_9(\Lambda_9) = 3\alpha_s(\Lambda_9)$ by integrating the one-loop RGE. In this scenario, depending on the size of $\alpha_9$ needed for instanton effects to be large enough, the UV unification scale $\Lambda_9$ need be no more than a loop factor above the flavor-breaking scale $\Lambda_3$. Suppressing the colored scalar masses relative to $\Lambda_9$ is then only a mild tuning. Of course one could also consider raising $\Lambda_3$, or the scalars getting an intermediate mass $\Lambda_3 < M < \Lambda_9$, or keeping fewer scalars light and having larger scale separation; we leave further consideration for future work aimed at more-realistic phenomenology.

\subsection{No Quality Problem} \label{app:quality}

Axion solutions to strong CP famously suffer a severe `quality problem' that their $U(1)_{\rm \scriptscriptstyle PQ}$ symmetries are easily destabilized. 
While it is true that we also require a good $U(1)_{\rm \scriptscriptstyle PQ}$ symmetry, our model is not plagued by a quality problem.
There are two issues to discuss. The first, purely field theoretic issue is how to clearly understand what it means to impose an anomalous symmetry on a classical action. While the program of generalized symmetries may lead us to better-understand the sense in which we can think of instanton effects are spurions for the anomalous symmetry, this issue is as yet not entirely clear.

A further and more general concern is that quantum gravitational effects are expected to violate any global symmetries, such that demanding an exact global symmetry in the UV seems theoretically unsound. We still have little understanding of precisely how such violation occurs, but one is at least motivated to consider the effects of Planckian operators made out of the fields in our theory which preserve only gauge symmetries and violate global ones. The potentially-disastrous implications of this for axion models were first discussed in \cite{Ghigna:1992iv,Barr:1992qq,Holman:1992us,Kamionkowski:1992mf}.

That this challenge is far worse for the axion solution than for the massless quark solution results from the following facts
\begin{enumerate}
 \item The presence of a scalar field $\phi$ charged under the $U(1)_{\rm \scriptscriptstyle PQ}$ symmetry allows global-symmetry-violating terms of any dimension e.g. $\mathcal{L} \supset c_n M_{\rm \scriptscriptstyle Pl}^{4-n} \phi^n$ which could conceivably be generated by quantum gravitational effects. 
 \item Astrophysical constraints on weakly coupled particles interacting with SM quarks impose a lower limit on the `axion decay constant', $f_a = \langle \phi \rangle$, $f_a \gtrsim 10^8 \text{ GeV}$ \cite{ParticleDataGroup:2020ssz}.
 \item The gravitational, global-symmetry violating operators must be subleading to the QCD-sized potential, which must localize $a \sim -\theta$ to very high accuracy. This means the fight is between $\Lambda_{\rm \scriptscriptstyle QCD}^4 \left(1 - \cos(a + \theta)\right)$ and $f_a^4 (f_a/M_{\rm \scriptscriptstyle Pl})^{n-4} \left(1 - \cos(n a + \varphi_n)\right)$, where $\varphi_n = \arg c_n$, and the minimum of $a+\theta$ must end up smaller than $\bar \theta$.
\end{enumerate}
Even granting some to-our-knowledge-unknown argument that the relevant operators are not present, if the dimension-5 operator has a random $\mathcal{O}(1)$ phase, its coefficient must satisfy
\begin{equation}
    |c_5| \lesssim 10^{-35} \left(\frac{\bar \theta}{10^{-10}}\right) \left(\frac{10^8 \text{ GeV}}{f_a}\right)^5,
\end{equation}
very clearly violating any naturalness principle and stretching the plausibility of these models unless there is extra structure that forbids these operators to high orders.

In contrast, the quality requirement for the massless quark solution is far less stringent. Beginning in the UV with a PQ symmetry at $\text{Re}(M) = \text{Im}(M) = 0$, instantons violate the PQ symmetry but not CP to produce an additive renormalization of $M$ in the direction $\text{Re}(M) >0$. A PQ-violating operator with an $\mathcal{O}(1)$ phase results in an additional contribution to $M$ in a general direction in the complex plane, but as long as the overall size of this contribution is small then the resulting $M$ is near the real axis and $\bar \theta \sim \text{Im}(M)/\text{Re}(M)$ stays small.

In the Standard Model massless up quark solution the leading dimensionful PQ-violating gauge-invariant operator is $|H|^2 \tilde{H} Q \bar u / M_{\rm \scriptscriptstyle Pl}^2$, and the effects of this operator must only compete with the dimensionless yukawa couplings. An additional such `bare' contribution to the up quark mass does not upset the massless up quark solution so long as $\text{Im}(m_u) \lesssim \bar \theta m_u$, which is easily satisfied as $v^2/M_{\rm \scriptscriptstyle Pl}^2 \sim 10^{-32}$. 

Now in comparison to the SM solution, which we know is not realized in nature, the model above necessarily needs new gauge dynamics at larger scales. This means there will be irrelevant operators with UV scalars which get larger vevs, for example $|\Phi|^2 \tilde{H} Q \bar d / M_{\rm \scriptscriptstyle Pl}^2$ or $\tilde{H} Q \Sigma \bar d/M_{\rm \scriptscriptstyle Pl}$. In the former case, these operators are completely safe so long as $\langle \Phi \rangle \sim \Lambda_9 \lesssim 10^{13} \text{ GeV}$, while in the latter case there are no naturalness concerns so long as $\langle \Sigma \rangle \sim \Lambda_3 \lesssim 10^8 \text{ GeV}$. Then there are no issues with the model of Section \ref{sec:fromUV} even if these PQ-violating operators are generated by quantum gravity with $O(1)$ coefficients and random phases. 

Overall rather than posing a problem, these naturalness considerations motivate the part of this model's parameter space in which the effects are most visible in low-energy experiments. That is, while generally $\Lambda_3$ could be at some high scale and the model still works great, the above quantum gravitational concerns suggest BSM quark flavor-changing physics should be, at worst, still in striking distance in the mid-to-long term \cite{EuropeanStrategyforParticlePhysicsPreparatoryGroup:2019qin}. 

\section{Further Issues and Directions} \label{sec:generalize}

In this work we have uncovered and explored a new feature of the colored Standard Model fermions. 
Remarkably, the known particle spectrum admits gauged flavor symmetries which bear out non-invertible symmetry structures through which ultraviolet instantons may resolve infrared naturalness issues.  
In the quark sector, this non-invertible symmetry appears when there is a non-trivial global structure for the gauge group---a possibility which is permitted because $N_c = N_g$. 
Following our infrared non-invertible symmetry analysis, we have sketched a scheme for implementing the massless down-type quarks solution to strong CP problem in $SU(9) \rightarrow SU(3)^2/\mathbb{Z}_3\rightarrow SU(3)_C$. There are many directions for further investigation, and we briefly discuss some of them in no particular order.

\paragraph{Alternative symmetry-breaking}
It is potentially appealing to skip the $SU(3)_H$ phase of the theory entirely, and go directly from $SU(9)$ to $(SU(3)_C \times U(1)_H) /\mathbb{Z}_3$ before breaking to $SU(3)_C$. 
From the UV, this route might allow us to avoid introducing the large Higgs representation 165 which is called for by $SU(9) \rightarrow SU(3)^2/\mathbb{Z}_3$, which would imply less suppression of the instanton density. 
From the IR, the abelian horizontal symmetry is constrained only to a few TeV, see \cite{Liu:2011dh,Crivellin:2015lwa} for constraints from the LHC and low-energy flavor, and \cite{Langacker:2000ju} for the general framework. An abelian horizontal phase then allows potentially more-visible observational signatures of such theories at the energy frontier.

\paragraph{Full flavor}
The model we have studied above not did introduce enough structure to generate a realistic flavor sector, and we can easily understand the sense in which we are missing ingredients. Keeping the Higgs $H$ as a color-flavor singlet means that the flavor-singlet coupling to the SM Higgs is allowed, and generically leads to $y_u, y_d \propto \mathds{1} + \text{perturbations}$. Unfortunately, the observed SM quark spectrum is just not so amenable to fitting this structure while keeping a good notion of power-counting for the perturbations.
Likely the needed solution is to embed the Higgs in a color-flavor adjoint. The SM Higgs then arises as a linear combination of the colorless, flavor adjoint $80 \supset (1,8)$, and the orientation of the Higgs in this matrix provides another handle for producing the SM yukawa structure. This sort of structure has recently been utilized in \cite{Davighi:2022fer} in a UV theory which is spiritually related to ours. 

\paragraph{Interplay with other strong CP strategies} 
We have utilized a mechanism of producing hermitian yukawas to ensure our UV boundary condition $\bar \theta(\Lambda_9)$ is not disrupted while generating the CKM matrix, but there may be other choices. The addition of vector-like quarks would allow much greater freedom in shaping the yukawas, but may require a Nelson-Barr-like structure to avoid introducing dangerous phases.
On the other hand our avoidance of vector-like quarks makes the scheme of Section \ref{sec:flavorBreaking} seem well-suited as a novel sort of model for spontaneous CP violation, which we will explore in the forthcoming \cite{Koren:2024c}.
There may also be a role for a parity symmetry in a left-right extension of this model where the broken $SU(2)_R$ is responsible for the initial distinction between up and down quarks.  Another obvious strategy we are currently pursuing \cite{Koren:2024b} is to spontaneously break the $U(1)_{\rm \scriptscriptstyle PQ}$ to find a model for a `heavy QCD axion' with additional mass provided by the small instantons, whose phenomenology has been much-discussed in recent years (see e.g. \cite{Holdom:1982ex,Holdom:1985vx,Rubakov:1997vp,Berezhiani:2000gh,Hook:2014cda,Fukuda:2015ana,Dimopoulos:2016lvn,Agrawal:2017ksf,Gaillard:2018xgk,Gherghetta:2016fhp,Gherghetta:2020keg,Kivel:2022emq,Valenti:2022tsc,Csaki:2019vte,Hook:2019qoh}) and whose domain walls have rich phenomenology due to the non-invertible symmetry \cite{Cordova:2023her}.

\paragraph{Relation to Agrawal \& Howe \cite{Agrawal:2017evu}}
The scenario of Agrawal \& Howe also utilizes small instantons to revive the massless quark solution to the strong CP problem, but not in a unified manner.
They employ a flavor deconstruction of $SU(3)_C \subset SU(3)_1 \times SU(3)_2 \times SU(3)_3$ and so begin with three free diagonal yukawas (top, charm, down), then at the breaking scale $\Lambda_3$ the instantons of their three different gauge groups generate the three other diagonal yukawas with appropriate sizes. Resultingly the only model-building they need do then is to generate the off-diagonal entries. 
Indeed, by cleverly adding a few fields with judiciously chosen charges, they are able to fit the full CKM structure of the SM including $\delta_{\rm \scriptscriptstyle CKM}$. They then find that this induces $\bar \theta$ as a 2-loop correction which depends on the small off-diagonal yukawas of the SM quarks, and so can be slightly below the current upper bound, $\Delta \bar \theta \lesssim 10^{-10}$, and within experimental reach.

We note that (at least structurally) one could attempt to embed the model of Agrawal \& Howe in the same UV theory we have, $SU(9) \supset SU(3)^3$. 
With the aesthetic and reductionist appeal of unification one would have additional challenges of then explaining how to end up with the appropriate flavor-asymmetric PQ symmetries, and the varying sizes of nonzero yukawas and gauge couplings $g_1 > g_2 > g_3$, rather than inputting these by hand. 
A particular difficulty in this case may be that the up-type/down-type hierarchy flips in the first generation relative to the others, and it seems non-trivial for such a PQ to appear starting with a flavor-unified theory.  
Still, it would be interesting to think more about a unified model embedding Agrawal \& Howe, which would mirror the strategy of Davighi \& Tooby-Smith \cite{Davighi:2022fer} implementing flavor deconstruction and reunification in the electroweak sector.
If such a scheme can work, this two-step breaking pattern from $SU(9)$ to $SU(3)_C$ has the advantage of generating the diagonal and off-diagonal structures at different scales, but misses out on having the instantons and flavor-breaking at different scales.  

\paragraph{Quark-lepton Color-flavor Unification}

We were led to this model by a parallel with our earlier work on neutrino masses protected by non-invertible symmetries \cite{Cordova:2022fhg}. 
A unification of these parallel stories generating both tiny Dirac neutrino masses and small strong CP violation may be achievable in a flavor-twisted Pati-Salam theory $SU(12)\times SU(2)_L \times SU(2)_R$, which we are currently investigating. 
Such non-trivial gauge-flavor unified theories have received astonishingly little attention \cite{Kuo:1984md,Davighi:2022fer,Davighi:2022qgb}, and are ripe for further phenomenological study.

Very generally, the discovery of such unified theories gives interesting top-down motivation to study separate gauged quark and lepton flavor symmetries. 
It is evident from experimental bounds that the scale at which non-abelian flavor physics is generated in the quark sector is far beyond the reach of the energy frontier. 
Na\"{i}vely one might conclude that if we ultimately want quark-lepton unification, this should imply that there is a single flavor symmetry acting on both, meaning that the non-abelian lepton flavor scale is also necessarily very high.
Not so, as breaking $SU(12) \rightarrow SU(9)_{\text{quarks}} \times SU(3)_{\text{leptons}} \times U(1)_{B-L}$ means that lepton flavor may yet be generated at accessible energy scales consistently with unification.
In particular this provides further UV motivation for a muon collider to explore the energy frontier of lepton flavor physics (see e.g. \cite{Dasgupta:2023zrh}).

\paragraph{Higgsed Lattice Field Theory} 
To fully understand the phenomenological potential of this model we would like to know the bottom/top yukawa ratio generated by the $SU(9)$ instantons over the entire parameter space. Since perturbative corrections are enhanced by $N$ in large $N$ gauge theories for a given gauge coupling $\alpha$, we quickly lose good control, and so numerical computations will be necessary to accurately understand the physics.
Lattice computations of Higgs-Yang-Mills theories have received relatively little attention recently, but for early discussions see e.g. \cite{Brower:1982yn,Lee:1985yi,Shrock:1987bx}.
Here we also have the challenge of trying to understand the size of the $SU(9)$ instanton effects without needing to specify a fully-realistic theory in which we could check literally that we land on the SM spectrum of hadrons. 
One could perhaps input a potential which only condenses the scalar field $\Phi$ effecting $SU(9) \rightarrow SU(3)^2/\mathbb{Z}_3$, letting the flavor group confine as well, and inferring the ratio of yukawas from the relative masses of the flavor-symmetric $\Delta^{++} = uct$ and $\Delta^{-} = dsb$. 

\section*{Acknowledgements}

We are grateful to T. Daniel Brennan, Richard Brower, Antonio Delgado, Samuel Homiller, Adam Martin, Kantaro Ohmori, Lian-Tao Wang for useful conversations, and thank Samuel Homiller and Adam Martin for comments on a draft of this work. 
Feynman diagrams have been created using FeynGame \cite{Harlander:2020cyh}.
We thank the Aspen Center for Physics (supported by a National Science Foundation grant
PHY-2210452) for the opportunity to participate in a summer workshop in 2023, during
which part of this work was completed. 
The work of CC is supported by DOE grant DE-SC0024367, by the Simons Collaboration on Global Categorical Symmetries, and by the Sloan Foundation. 
The work of SH is supported by the National Research Foundation of Korea (NRF) Grant RS-2023-00211732, by the Samsung Science and Technology Foundation under Project Number SSTF-BA2302-05, and by the POSCO Science Fellowship of POSCO TJ Park Foundation. The work of SK has been supported by an Oehme
Postdoctoral Fellowship and an Arthur H. Compton Lectureship from the EFI at UChicago, and by the NSF grant PHY-2112540. 
SK additionally thanks the Simons Center for Geometry and Physics at Stony Brook University, the Mainz Institute for Theoretical Physics (MITP) of the Cluster of Excellence PRISMA+ (Project ID 39083149), and the Korea Advanced Institute of Science and Technology for hospitality during the completion of this work.

\appendix

\section{Fractional and CFU Instantons}
\label{app:fractionalInstanton}

In this section, we review a few facts about fractional instantons. Non-invertible symmetries appearing in this work are connected to a certain kind of a fractional instanton configuration which may be loosely thought of as a special linear combination of fractional instantons of more than one gauge group factors, e.g.~$SU(3)_C \times SU(3)_H$. These are examples of ``Color-Flavor-$U(1)$ (CFU) instantons introduced in \cite{Anber:2021iip} (see also \cite{Anber:2019nze}). The discussion presented here follows closely \cite{Brennan:2023mmt, Anber:2021iip}.~\footnote{For more reviews on generalized symmetries, see \cite{Gomes:2023ahz, Bhardwaj:2023kri, Luo:2023ive, Shao:2023gho} and the appendices of \cite{Brennan:2023kpw} for a brief review.}

\subsection{Fractional Instantons}

The simplest example of a gauge theory with fractional instantons is $PSU(N) = SU(N) / \mathbb{Z}_N$ gauge theory and we will focus on this.
Let us first consider an $SU(N)$ gauge theory with only matter fields in the adjoint representation. In this case, the entire set of fields, including the gauge fields, are invariant under $\mathbb{Z}_N$ center transformations. This in turn means that there are $\mathbb{Z}_N$-charged Wilson lines that cannot be screened since there exists no charged particle that can cut those lines: such a theory enjoys 1-form electric $\mathbb{Z}_N^{(1)}$ symmetry. Since it is a good quantum symmetry, it (or any subgroup of it) can be gauged. If we gauge the entire $\mathbb{Z}_N^{(1)}$ electric center, the resulting theory is $SU(N) / \mathbb{Z}_N$ gauge theory. The latter has no more 1-form electric symmetry but this time it has 1-form dual magnetic $\mathbb{Z}_N^{(1)}$ center symmetry. This fact can also be checked from the fact that $\pi_1 (SU(N) / \mathbb{Z}_N) = \mathbb{Z}_N$. As we describe now, this means that the path integral of $SU(N) / \mathbb{Z}_N$ gauge theory includes a summation over dynamical 2-form $\mathbb{Z}_N$ gauge fields $B_2$ which turn on $\mathbb{Z}_N$-valued magnetic flux. 
\beq
\oint_{\Sigma_2} \frac{B_2}{2\pi} = \frac{1}{N} \mathbb{Z}.
\eeq
And the theory contains fractional instanton configurations as well as regular integer valued instantons. 

To see this more explicitly, we start with $U(N) = \frac{SU(N) \times U(1)}{\mathbb{Z}_N}$ gauge theory and reduce it down to $SU(N) / \mathbb{Z}_N$. We do this in two steps. First, we project down the local degree of freedom associated with $U(1)$ factor. We can do this by adding a Lagrange multiplier term to the action of the $U(N)$ theory.
\beq
S = \frac{1}{g^2} \int \text{Tr} \left( \hat{f}_2 \wedge * \hat{f}_2 \right) + \frac{i}{2\pi} \int F_2 \wedge \text{Tr} \left( \hat{f}_2 \right) + \frac{i \theta}{8\pi^2} \int \text{Tr} \left( \hat{f}_2 \wedge \hat{f}_2 \right).
\label{eq:PSUN_action_1}
\eeq
Here, $\hat{f}_2$ is the 2-form field strength of the $U(N)$ gauge field $\hat{a}_1$ and $F_2$ is a 2-form Lagrange multiplier whose equation of motion imposes $\text{Tr} \left( \hat{f}_2 \right) = 0$ which projects out the $U(1)$ degree of freedom. This theory is just SU(N) and so still has $\mathbb{Z}$-valued instanton spectrum. To achieve $SU(N) / \mathbb{Z}_N$ theory with fractional instantons, we impose 1-form $\mathbb{Z}_N^{(1)}$ symmetry. To this end, we introduce $SU(N)$ gauge field $a_1$ and its field strength $f_2$. These are related to $U(N)$ fields as
\beq 
\hat{a}_1 = a_1 + \frac{1}{N} A_1 \mathds{1}, \;\;\;\; \hat{f}_2 = f_2 + \frac{1}{N} d A_1 \mathds{1}
\eeq 
where $A_1$ is a $U(1)$ gauge field and $\mathds{1}$ is the $N \times N$ identity matrix. 
Next, we impose $\mathbb{Z}_N^{(1)}$ symmetry under which fields transform as 
\bea
a_1 \to a_1 , \;\; A_1 \to A_1 - N \lambda_1 \;\; \Rightarrow \;\; \hat{a}_1 \to \hat{a}_1 - \lambda_1 \mathds{1}, \; \hat{f}_2 \to \hat{f}_2 - d \lambda_1 \mathds{1}
\eea 
where $\lambda_1$ is the 1-form symmetry transformation parameter. One then realizes that the action \eqref{eq:PSUN_action_1} is not invariant under the above symmetry transformation. It can be made invariant by introducing a 2-form field $B_2$ and assigning a symmetry transformation rule $B_2 \to B_2 - d \lambda_1$. Then, invariant action is given by
\bea 
S  = && \frac{1}{g^2} \int \text{Tr} \left[ \left( \hat{f}_2 - B_2 \mathds{1} \right) \wedge * \left( \hat{f}_2 - B_2 \mathds{1} \right) \right] + \frac{i}{2\pi} \int F_2 \wedge \text{Tr} \left( \hat{f}_2 - B_2 \mathds{1} \right) \nonumber \\ 
&& + \frac{i \theta}{8\pi^2} \int \text{Tr} \left[ \left( \hat{f}_2 - B_2 \mathds{1} \right) \wedge \left( \hat{f}_2 - B_2 \mathds{1} \right) \right].
\label{eq:PSUN_action_2_B2}
\eea
The appearance of the combination $\left( \hat{f}_2 - B_2 \mathds{1} \right)$ can be thought of as the coupling the theory to the 2-form gauge field $B_2$ of the $\mathbb{Z}_N^{(1)}$ electric symmetry.
To understand the instanton spectrum, we first note that $\text{Tr} \left(\hat{f}_2 \right) = d A_1$. Then the Lagrange multiplier term turns into
\beq 
\frac{i}{2\pi} \int F_2 \wedge \left( d A_1 - N B_2 \right) 
\eeq
which describes $\mathbb{Z}_N$ gauge theory (often called a BF theory). Specifically, the term $\frac{iN}{2\pi} F_2 \wedge B_2$ corresponds to the usual Lagrangian for the $\mathbb{Z}_N$ BF theory and the above action is obtained by dualizing the 1-form gauge field in $F_2 = d \tilde{A}_1$ to $A_1$. One may interpret \eqref{eq:PSUN_action_2_B2} as $SU(N)$ theory coupled to this $\mathbb{Z}_N$ BF theory \cite{Kapustin:2014gua}. 
A brief review on BF theory can be found in the appendix B of \cite{Brennan:2023kpw}, and for more detailed discussions we refer to \cite{Banks:2010zn, Kapustin:2014gua, Brennan:2023mmt}. For instance, the equation of motion for $F_2$ sets $B_2 = d A_1 / N$, showing that $B_2$ is a $\mathbb{Z}_N$ gauge field.

Finally, the $\theta$-angle term is given by
\bea
S_\theta &&= \frac{i \theta}{8\pi^2} \int \text{Tr} \left( \hat{f}_2 \wedge \hat{f}_2 \right) - N B_2 \wedge B_2 \nonumber \\
&& = \frac{i \theta}{8\pi^2} \int \text{Tr} \left( \hat{f}_2 \wedge \hat{f}_2 \right) - \text{Tr} \left( \hat{f}_2 \right) \wedge \text{Tr} \left( \hat{f}_2 \right) + \frac{i \theta}{8\pi^2} \int N (N-1) B_2 \wedge B_2 \nonumber \\
&& = i \theta n + i \theta \left( \frac{N-1}{2 N} \right) \int w_2 \wedge w_2.
\label{eq:PSUN_fractional_inst}
\eea 
In the second line, we subtracted and added the combination $\text{Tr} \left( \hat{f}_2 \right) \wedge \text{Tr} \left( \hat{f}_2 \right)$ so that the first integral there becomes the standard integer valued $SU(N)$ instanton number $n$, while the second term captures the fractional instanton effects. To get the last line, we defined $w_2 = N \frac{B_2}{2\pi}$ (called the second Stiefel-Whitney class) whose integral is integer valued 
\beq 
\oint w_2 = 0, 1, \cdots, (N-1).
\eeq 
In a theory where fermions can be introduced (technically, a spin-structure can be defined), 
$\int w_2 \wedge w_2 \in 2 \mathbb{Z}$ and this then clearly shows that the second term in the last line of \eqref{eq:PSUN_fractional_inst} indeed corresponds to the fractional instanton.

\subsection{CFU Instantons}
\label{app:CFU}

Having described the fractional instantons of $SU(N)/\mathbb{Z}_N$ theory, we now briefly discuss CFU (Color-(non-abelian)Flavor-$U(1)$) instantons \cite{Anber:2021iip}. The original construction given in \cite{Anber:2021iip} was the most general and refined background fields for the 1-form electric center symmetry of $SU(N)$ gauge theory. The novelty is that given fermion contents with non-abelian flavor symmetry say $SU(F)$ and a global $U(1)$ symmetry, CFU background analysis yields the most general electric 1-form symmetry of the theory. This then allows to determine the largest set of mixed 0-form and 1-form 't Hooft anomalies, hence the strongest constraints on the IR phases of the strongly coupled gauge theory.   

Let us denote the set of matter contents as $\lbrace \Psi_i \rbrace$. Also, let us write $G_c, G_f, G_u$ for color, non-abelian flavor and $U(1)$ groups. However, the following analysis can be generalized to any choice and any number of groups. Then, we write 0-form center transformations acting on $\Psi_i$'s as $z_c \in Z (G_c), \; z_f \in Z (G_f)$ and $z_u \in Z (G_u)$. Calling Wilson lines of $G_c, G_f, G_u$ as $W_c, W_f$ and $W_u$, the 1-form electric symmetry can be figured out by understanding the spectrum of topologically protected (i.e.~unscreened by local charges) Wilson lines. This includes all possible composite Wilson lines of the form $W_c^\ell W_f^m W_u^n, \; \ell, m , n \in \mathbb{Z}$. Protected Wilson line spectrum is obtained by solving so called `cocycle conditions'. The basic idea is to determine the most general set of center transformations that act trivially on the \emph{entire} $\lbrace \Psi_i \rbrace$ (and gauge fields)
\beq 
\Psi_i \to z_c z_f z_u \Psi_i = \Psi_i \;\;\; \forall i .
\eeq 
Since the entire fields of the theory are uncharged under any such center transformations, associated (composite) Wilson lines are not screened, and therefore electric 1-form symmetry is determined. Focusing on the case $G_c = SU(N), G_f = SU(F)$ and $G_u = U(1)$, the solutions take the form 
\beq
z_c = e^{\frac{2\pi i}{N} \ell}, \;\; z_f = e^{\frac{2\pi i }{F} m}, \;\; z_u = e^{2\pi i n},
\eeq
with $\ell = 0, \cdots, (N-1), \; m = 0, \cdots, (F-1), \; n \in [0,1]$. Then, it can be shown that a particular combination (determined by $\{ \ell, m, n \}$) of 2-form background gauge fields, call them $B_c, B_f, B_u$, of individual 1-form electric center symmetry can be consistently activated. If no solution with only a single center factor exists, e.g.~$(\ell\neq 0 , m=0, n=0)$, 1-form electric symmetry is only defined in terms of composite Wilson operators and in that case one has to turn on a specific combination of 2-form background fields controlled by the solution $\{ \ell, m, n \}$ as an acceptable configuration. 

Non-vanishing 2-form background gauge fields lead to CFU instanton (or CFU topological charge). Using the expression of fractional instanton \eqref{eq:PSUN_fractional_inst}, it is given by the combination of the following three (C, F, and U) fractional instantons
\bea
&& \mathcal{N}_c = \left( \frac{N-1}{N} \right) \int \frac{w_c \wedge w_c}{2} = \ell_1 \ell_2 \left( 1 - \frac{1}{N} \right) \\
&& \mathcal{N}_f = \left( \frac{F-1}{F} \right) \int \frac{w_f \wedge w_f}{2} = m_1 m_2 \left( 1 - \frac{1}{F} \right)  \\
&& \mathcal{N}_u = \int \frac{B_u \wedge B_u}{8\pi^2} = n_1 n_2 
\eea 
where $w$-fields are defined in terms of $B$-fields as in the previous section (see below \eqref{eq:PSUN_fractional_inst}). Also, when $\{ \ell_1, m_1, n_1 \} \neq \{ \ell_2, m_2, n_2 \}$, the above expression means the following. Consider the spacetime manifold of the form $M_4 = \mathbb{S}^2 \times \mathbb{S}^2$. Then, given two solutions $\{ \ell_1, m_1, n_1 \}$ and $\{ \ell_2, m_2, n_2 \}$ to the cocycle conditions, we take a background configuration which is a formal sum of $\{ \ell_1, m_1, n_1 \}$ units of CFU-fluxes piercing the first $\mathbb{S}^2$ with zero flux through the second $\mathbb{S}^2$ and configuration with zero flux through the first $\mathbb{S}^2$ and $\{ \ell_2, m_2, n_2 \}$ units of CFU-fluxes through the second $\mathbb{S}^2$. This leads, for the color part, to 
\beq
\oint_{\mathbb{S}^2 \times \mathbb{S}^2} \frac{w_c \wedge w_c}{2} = \ell_1 \ell_2,
\eeq
and similarly for the flavor and $U(1)$ parts.
Finally, we stress again that $\mathcal{N}_c, \mathcal{N}_f, \mathcal{N}_u$ separately are not well-defined configurations. Only the whole combination makes sense and gives rise to integer-valued (i.e.~sensible) Dirac indices. When the electric 1-form symmetries are gauged, then the background fields $B_{c,f,u}$ are dynamical fields and path integrated, and the CFU configurations turn into dynamical instantons of the theory. See \cite{Anber:2021iip} for more details and usage of CFU instantons. 

\section{Global Structure and Non-Invertible Symmetries} \label{app:globalStructure}

\subsection{SM Global Structure}

In Section \ref{sec:fromIR} we learned that the non-invertible symmetries of $\suto$ or $\sutt$ depend sensitively on the global structure, and the effects of interest are not present in the absence of this `modding' of the gauge group. 
With the goal of providing additional background to readers unfamiliar with these concepts, in this appendix we review the global structure of the SM itself, $G_{\text{SM}} = (SU(3)_C \times SU(2)_L \times U(1)_Y) / \Gamma$ and conclude that the SM non-invertible symmetries do not depend on the choice $\Gamma \in \lbrace \mathds{1}, \mathbb{Z}_2, \mathbb{Z}_3, \mathbb{Z}_6\rbrace$. 
We refer to  \cite{Tong:2017oea} for a discussion of line operator spectrum (Wilson, 't Hooft and Dyonic) depending on the SM global structure, to \cite{Anber:2021upc} for the (fractional) instanton spectrum and related cosmological impacts, and to \cite{Cordova:2023her} for the discussion of SM global structure and axion non-invertible symmetry and resulting observational implications in terms of axion domain wall physics. 

In Table \ref{tab:charges} we show the representations of the SM fermions under the gauge symmetries and classical global symmetries of the SM, in addition to the right-handed neutrinos and the Higgs boson. One may observe that there are certain combinations of center transformations which act trivially on all of the fields of the SM. For example, the center of $SU(2)_L$ is a $\mathbb{Z}_2$ subgroup whose nontrivial element acts as $(-1)\mathds{1}_2$ on the fields charged under $SU(2)_L$. These effects can be compensated exactly by a rotation by $\pi$ of hypercharge, the $\mathbb{Z}_2 \subset U(1)_Y$ subgroup, so that if we do a diagonal transformation by $(-\mathds{1}_2) e^{i \pi Y}$, then e.g. 
\begin{equation}
    Q_i \rightarrow Q_i (-1) e^{i \pi (+1)} = Q_i, \quad L_i \rightarrow L_i (-1) e^{i \pi (-3)} = L_i, \quad \bar u_i \rightarrow \bar u_i e^{i \pi (-4)} = \bar u_i, \quad \dots
\end{equation}
and all the SM fields are invariant under such a rotation (recall that the $W$ gauge fields are in the adjoint, two-index representation of $SU(2)$ so transform as $(-1)^2$). The fact is that the observed fields with odd numbers of $SU(2)$ indices all also have odd hypercharge, while those with even numbers of $SU(2)$ indices all have even hypercharge. And it is not difficult to see that the quarks have just the right hypercharges that a similar diagonal $\mathbb{Z}_3$ subgroup of $SU(3)_C \times U(1)_Y$ also acts trivially on the known fields.

In fact, it is not too difficult to show that the entire SM fields, matter as well as gauge fields, are invariant under $\mathbb{Z}_6$ center transformations. To show this explicitly, we first note that the centers of $SU(3)_C, SU(2)_L$ and $\mathbb{Z}_6$-center of $U(1)_Y$ are generated by
\beq
e^{2\pi i \lambda_8 / 3} = e^{2\pi i /3} \mathds{1}_3 \in SU(3)_C \hspace{0.5cm} e^{2\pi i T_3 / 2} = - \mathds{1}_2 \in SU(2)_L \hspace{0.5cm} e^{2\pi i q_Y/6} \in U(1)_Y
\eeq 
where $\lambda_8 = {\rm diag} (1,1,-2)$ and $T_3 = {\rm diag} (1,-1)$, and $q_Y$ denotes the generator of $U(1)_Y$ normalized as $e^{2\pi i q_Y}=1$. Then, the statement is that the representations of SM fields are such that the combination
\beq
e^{2\pi i \lambda_8 / 3} \times e^{2\pi i T_3 / 2} \times e^{2\pi i q_Y/6} 
\label{eq:Z_6_generator}
\eeq
acts trivially (i.e.~as an identity) on all fields of the SM. Realizing \eqref{eq:Z_6_generator} as the generator of $\mathbb{Z}_6$ proves that the whole SM is uncharged under $\mathbb{Z}_6$. 

In terms of 1-form symmetry and related global structure, this means the following. Let us denote a general Wilson line of the SM as $W_C^a W_L^b W_Y^c$ where $W_C, W_L, W_Y$ are charge-1 Wilson line of $SU(3)_C, SU(2)_L$ and $U(1)_Y$, respectively, and $a,b,c \in \mathbb{Z}$. This particular Wilson line can be thought of as the worldline of a probe particle whose representation is such that it carries $N$-ality (roughly the number of boxes of Young tableau) of $(a,b,c)$ under the Lie algebra $su(3)_C \times su(2)_L \times u(1)_Y$. If there exists a dynamical/light particle with $N$-ality a divisor of $(a,b,c)$, then a pair-production can cut the Wilson line and we say the corresponding line is screened. In the absence of dynamical charges to screen, on the other hand, the Wilson line is stable and we have a moment to discuss 1-form global symmetry. In particular, such a topologically protected Wilson line shows the existence of electric 1-form center symmetry. Representations of SM is then consistent with the existence of $\mathbb{Z}_6$ stable Wilson lines and $\mathbb{Z}_6^{(1)}$ 1-form electric symmetry. This is the case when the global structure is $\Gamma = \mathds{1}$.

Non-trivial $\Gamma = \mathbb{Z}_2, \mathbb{Z}_3, \mathbb{Z}_6$ is obtained if electric $\mathbb{Z}_6^{(1)}$ or its subgroup is gauged. This means that the 2-form background gauge field of the 1-form electric center, denoted before $B_2$, is now path integrated. If we gauge $\mathbb{Z}_q \subset \mathbb{Z}_6, \; q=2,3,6$, we get $\Gamma = \mathbb{Z}_q$. Gauging an electric symmetry generates the dual magnetic symmetry \cite{Tachikawa:2017gyf}. Indeed, with $\mathbb{Z}_q \subset \mathbb{Z}_6$ gauged, SM now has $\mathbb{Z}_{6/q}^{(1)} ({\rm e}) \times \mathbb{Z}_q^{(1)} ({\rm m})$, where the first (second) is the electric (magnetic) 1-form symmetry.

\subsection{Absence of Non-invertible Symmetries of SM}
\label{app:no_NIS_SM}

In this appendix, we show that with the up yukawas $y_u$ turned on, the SM does not possess any non-invertible symmetries acting on quark fields for all choices of global structure $\Gamma$. 

To this end, we recall that non-invertible symmetry can exist if there is global $U(1)$ which is primarily broken by $U(1)$ or fractional instantons. In this sense, it is already clear from Table~\ref{tab:IRanomaly} that non-invertible symmetry is not present in SM with any choice of $\Gamma \in \{ \mathbb{Z}_6, \mathbb{Z}_3, \mathbb{Z}_2 \}$: ABJ anomaly coefficients are flavor-universal and crucially $SU(3)_C$ instanton already achieves maximal breaking of all $U(1)$ symmetries of the theory. 
While fractional instantons, in general, lead to smaller anomaly coefficients, hence more breaking of $U(1)$ symmetry, the above fact clearly shows that the best $\mathbb{Z}_6$ fractional instantons of $SU(3)_C \times SU(2)_L \times U(1)_Y /\mathbb{Z}_6$ can do is to break $U(1)$ symmetries as much as regular $SU(3)_C$ instantons do. What is left is to demonstrate that this is indeed the case. 

A $\mathbb{Z}_6$ fractional instanton configuration can be constructed following Appendix~\ref{app:fractionalInstanton}. $\mathbb{Z}_6$ fractional instanton is given by a combination of $\mathbb{Z}_3$ instanton of $SU(3)_C / \mathbb{Z}_3$, $\mathbb{Z}_2$ instanton of $SU(2)_L / \mathbb{Z}_2$ and instanton of $U(1)_Y / \mathbb{Z}_6$. Following the notation defined in Appendix~\ref{app:fractionalInstanton}, we can write topological charges as
\begin{equation}
    \mathcal{N}_C = \ell_1 \ell_2 \left( 1 - \frac{1}{N_c} \right),~~~\mathcal{N}_L = m_1 m_2 \left( 1 - \frac{1}{2} \right),~~~\mathcal{N}_Y = n_1 n_2.
\end{equation}
A consistent and the minimal $\mathbb{Z}_6$ fractional instanton can be found by solving cocycle conditions \cite{Anber:2021iip}. The result is the combination of $\mathcal{N}_C, \mathcal{N}_L$ and $\mathcal{N}_Y$ with $\ell_1 = \ell_2 = -1$, $m_1 = m_2 = 1$ and $n_1 = n_2 = - \frac{1}{6}$. Calling this configuration $[\text{CLY}]$, topological charges take values $\mathcal{N}_C = \frac{2}{3},~ \mathcal{N}_L = \frac{1}{2},~ \mathcal{N}_Y = \frac{1}{36}$. The ABJ anomalies of $U(1)_{\tilde{B}_i}$ and $U(1)_{\bar{d}_i}$ are independent of the flavor index $i$ and are computed to be
\beq  
 \tilde{B}_i [\text{CLY}] = 1, \;\;\;\; \bar{d}_i [\text{CLY}] = 1.
\eeq 
We conclude that $\mathbb{Z}_6$ fractional instantons break $U(1)$ symmetries $\Pi_i U(1)_{\tilde{B}_i} \times U(1)_{\bar{d}_i}$ of the SM with $y_d = 0$ just as much as non-abelian $SU(3)_C$ instantons do, therefore no non-invertible symmetry exists with any choice of $\Gamma$.

\section{'t Hooft Vertices in \texorpdfstring{$SU(9)$}{SU(9)}} \label{app:SU9thooft}

In the quantum theory of $SU(9)$, nonperturbative gauge-theoretic effects generate 't Hooft vertices which dynamically violate the anomalous quantum numbers. As depicted in Figure \ref{fig:SU9Inst}, this results in the generation of the down-type yukawa, which in a general basis is parametrically
\begin{equation} \label{eqn:ybGenRough}
    \mathcal{L} \sim y_t^\star e^{i \theta_9} e^{-S_\text{inst}} H \mathbf{Q} \mathbf{\bar d},
\end{equation}
where with $S_\text{inst} = \frac{8\pi^2}{g_9^2(\Lambda_9)}$ we get the leading estimate for the instanton effects. Note that the instanton-induced down-type yukawa is $y_b \propto y_t^\star$. This ensures that the effective $\bar{\theta}$ still exactly vanishes: using chiral rotations, one may move phases in $y_t$ and $y_b$ into the $\theta$-term, and they exactly cancel. This is just an explicit manifestation of the observation made in \cite{Choi:1988sy}.

These 't Hooft vertices are precisely analogous to those which generate Dirac neutrino masses in the lepton $SU(3)_H$ theory of \cite{Cordova:2022fhg}. However, our use here is to generate the bottom yukawa---which is not parametrically small---$y_b / y_t \sim 1/40$, so this requires a more-precise understanding of the instanton-induced interaction. In particular we must include the polynomial factor which arises from quadratic fluctuations around the instanton background. 
We will ultimately find that in the parameter space where large $y_b$ could be generated, the one-instanton computation is not fully trustworthy, and the honest viability of this scenario will need to be determined by lattice simulations. So we aim at demonstrating the plausibility of generating $y_b$ of the right order of magnitude within this one-instanton approximation, and discuss the features of the quadratic fluctuations which enter the result only to this level.

The one-loop instanton gas computation was first performed in the pioneering work \cite{tHooft:1976snw} by 't Hooft  and for our usage consists essentially of calculating
\begin{equation}
    \left\langle H(x_1) \mathbf{Q}(x_2)\mathbf{\bar d}(x_3) \right\rangle_{\text{1-instanton}},
\end{equation}
which is the three-point correlator summed over all possible 1-instanton backgrounds, and then performing LSZ reduction to get the three-point amplitude. As effective field theorists we can think of finding the Wilsonian action obtained by integrating out instantons at smaller sizes $\rho$ ($\rho=$ the size modulus of the instanton) by matching to the amplitude $\mathcal{M}(H,\mathbf{Q},\mathbf{\bar d})$. Eventually we will integrate out `all' the instanton effects at the scale $\rho \sim \Lambda_9^{-1}$.
To justify this procedure in a Higgsed gauge theory, recall that in the case of a 1-instanton background, the action for the gauge field at the scale $\rho^{-1}$ is given by 
\begin{equation}\label{eq:instanton_gauge action}
    e^{-8\pi^2 / g^2 (\rho)} = e^{-8\pi^2 / g^2 (\Lambda_{\rm UV})} \left( \rho \Lambda_{\rm UV} \right)^{b_0} ,
\end{equation}
with $\Lambda_{\rm UV}$ the UV cutoff scale and $b_0>0$ the 1-loop $\beta$-function coefficient. However, the contribution from the scalar that achieves spontaneous breaking of the gauge group at $\Lambda_9$ behaves as 
\begin{equation}\label{eq:instanton_scalar action}
    e^{-8\pi^2 \rho^2 \Lambda_9^2},
\end{equation}
and we see that large instantons are exponentially suppressed \cite{Affleck:1980mp,Csaki:1998vv,Csaki:2019vte}. On the other hand, from \eqref{eq:instanton_gauge action} we see that the gauge contribution becomes more important for larger $\rho$ since the gauge coupling increases. In fact, the balance between the two contributions show that the dominant contribution comes from $\rho^2 \approx b_0 / 16\pi^2 \Lambda_9^2$, i.e. roughly $\rho \sim 1/\Lambda_9$, as can be studied systematically in the formalism of `constrained instantons' \cite{Affleck:1980mp}.

The correlator we want to calculate can be computed by a path integral 
\begin{equation}
    \int \mathcal{D}A \mathcal{D} \phi_i \mathcal{D}\psi_i \ H \mathbf{Q} \mathbf{\bar d} \ e^{-S_\text{gauge} - \int \mathcal{L}_\text{int}},
\end{equation}
integrating over the gauge field $A$ in the sector $\int F \tilde{F} /(32 \pi^2) = 1$ and also any charged scalars $\phi$ or fermions $\psi$. For detailed accounting of such integrals we refer to, for example \cite{Csaki:2019vte,Vandoren:2008xg,Morrissey:2005uza,Ringwald:1989ee,Espinosa:1989qn}, and we give here solely an overview. There are a number of sorts of effects which appear in this path integration, starting with the evaluation of bosonic modes
\begin{itemize}
    \item \textbf{Instanton background action}---The semi-classical gauge field background contributes to the action as
    \begin{equation}
        S = \int \frac{1}{4g^2} F^2 = \int \frac{1}{8g^2} ( F \pm \tilde{F})^2 \mp \int \frac{1}{4g^2} F \tilde{F},
    \end{equation}
    having made use of $F^2 = \tilde{F}^2$ and completing the square. Now we see that in a non-trivial topological sector with $\int F \tilde{F} /(32 \pi^2) = k$, $k \in \mathbb{Z}$, the action has a minimum at $S_{\text{1-inst}}=8\pi^2 |k|/g^2$ for (anti)self-dual field configurations $F = \pm \tilde{F}$. Thus the effects of the non-trivial saddles are suppressed by $\exp(-2\pi/\alpha)$, with $\alpha \equiv g^2/4\pi$.

    In a general basis where the UV theory contains also a $\theta$-term, $\mathcal{L} \supset \frac{i \theta}{32\pi^2} F \tilde{F}$ this phase enters the correlation functions in the instanton background just by evaluating this boundary term on the background solution. This provides the $\theta$-angle in \eqref{eqn:ybGenRough} in exactly the right combination to preserve $\bar \theta = 0$.
    
    \item \textbf{Gauge field zero modes}---the classical 1-instanton background, using the 't Hooft symbols $\eta_{a\mu\nu}$ and written in the `singular gauge', 
    \begin{equation}
        A_\mu(x) = \frac{2 \rho^2}{(x-x_0)^2} \frac{\eta_{a\mu\nu} (x-x_0)^\nu J^a}{(x-x_0)^2 + \rho^2}
    \end{equation} 
    possesses $4$ translational zero-modes corresponding to $x_0$, and $4N-5$ orientational zero-modes corresponding to the $SU(2)\subset SU(N)$ subgroup generators $J^a$. The final modulus is the size $\rho$ which is quantum mechanically non-trivial because the coupling evolves with scale. These zero modes are dealt with by the method of collective coordinates, and the proper normalization of these integrals leads to factors of $2^{-2N} (2 \pi/\alpha(\rho))^{2N}$. The integration over gauge rotations yields a further factor proportional $1/(N-1)!(N-2)!$ \cite{Bernard:1979qt}.
    \item \textbf{Charged field non-zero modes}---At the level of quadratic fluctuations about the instanton background, the matter field energy levels are skewed slightly away from their vacuum values. 't Hooft showed the determinants take the same form regardless of the statistics of the field, and the result depends solely on the degrees of freedom in different representations of the subgroup in which the instanton resides. 
    In total there is a factor $\exp\left[-(-1)^F \sum a(t) n(t)\right]$, where $a(t)$ is a certain polynomial 't Hooft derived which depends on the `isospin' $t$ of a given representation (taking the values $a(0) = 0, a(\half)=0.146, a(1)=0.443, a(\frac{3}{2})=0.853$), $n(t)$ counts the number of such representations, and fermions contribute inversely because of Grassmann statistics.
    The adjoint gauge field contributes $n(1) = 1$ and $n(\half) = 2(N-2)$, while the SM fundamental matter counts as $(-1)^F n(\half) = -4$.
\end{itemize}
Then after having integrated over all the bosonic modes except for the scaling modulus (and also the nonzero fermionic modes), the instanton density in an $SU(N)$ gauge theory takes the form 
\begin{equation} \label{eqn:instDens}
    D(\rho) = \frac{4}{\pi^2} \frac{2^{-2N} e^{-\sum (-1)^F a(t) n(t)}}{(N-1)!(N-2)!} \left(\frac{2 \pi}{\alpha(\rho)}\right)^{2N} \exp\left(-\frac{2\pi}{\alpha(\rho)}\right), 
\end{equation}
where $\rho$ is the size parameter of the instanton solution. 

\paragraph{\textbf{Charged fermion zero-modes}} Now for charged fermions there is a qualitatively new sort of effect. In an instanton background, charged fermions will have zero modes which are exactly counted by the Dirac index appearing in the coefficient of chiral anomalies. The path integral includes integrals over these zero-modes, which for our content of solely fundamental fermions implies one zero-mode integral per field
\begin{equation}
    \int \mathcal{D}\psi_i \supset \int \prod\limits_{i=1}^{N_f} d\xi^{(0)}_i,
\end{equation}
where $\xi^{(0)}_i$ is the Grassmann coefficient of the zero-mode eigenspinor of the $i^\text{th}$ field. For a Grassmann variable $\xi$, integration works as $\int d \xi = 0, \int d\xi \xi = 1$, so the presence of these Grassmann zero-mode integrals implies that correlation functions in the instanton background all vanish unless they include fermion fields to be integrated against those zero-modes. With just the SM fields, this means that the 1-instanton path integral immediately generates a 't Hooft vertex which contributes to correlation functions like 
\begin{equation}
    \left\langle \mathbf{Q} \mathbf{\bar d} \mathbf{Q} \mathbf{\bar u} \right\rangle.
\end{equation}
Instead, to get to the yukawa operator of interest this implies we need an insertion of a Lagrangian operator, meaning the nonvanishing contribution appears in the path integral as 
\begin{equation}
    \int \mathcal{D}(\text{fields}) \ H(x_1) \mathbf{Q}(x_2) \mathbf{\bar d}(x_3) \left(\int d^4y \ i y_t^\star \tilde{H}^\dagger(y) \mathbf{Q}^\dagger(y) \mathbf{\bar u}^\dagger(y) \right) e^{- \int \mathcal{L}}
\end{equation}
corresponding to the vertex insertion depicted in Figure \ref{fig:SU9Inst}. Despite what may look like loops of fermionic lines in the figure, there is no undetermined loop momentum there to integrate over---rather we integrate against the zero-mode wavefunction which is provided by the 't Hooft vertex, and this does not result in any loop suppression.

\begin{figure}
  \centering
  \includegraphics[width=.6\linewidth]{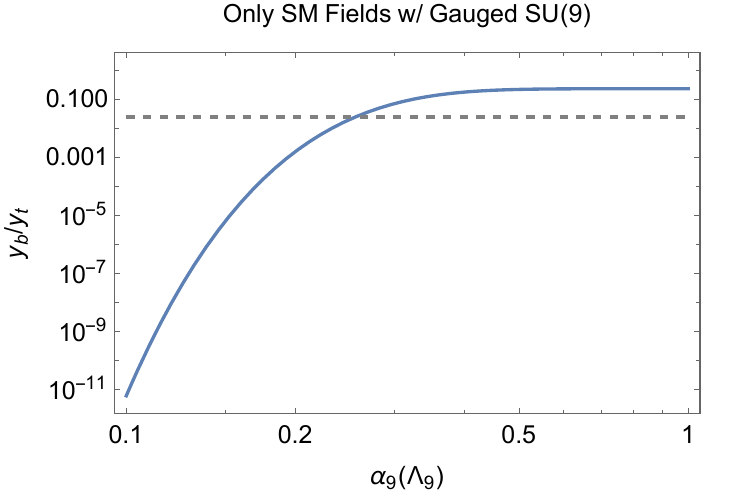}\hspace{.5cm}
  \caption{Size of $y_b/y_t$ which is generated by $SU(9)$ instantons at $\Lambda_9$, performing the integral in \eqref{eqn:inducedDownYuk} using the instanton density of \eqref{eqn:instDens} in the theory of solely the SM matter fields.  This theory is not realistic because it does not yet incorporate matter involved in breaking $SU(9) \rightarrow SU(3)_C$. The dashed line is the observed infrared value of $y_b/y_t$.}
  \label{fig:SU9InstSM}
\end{figure}

Past the SM matter and the $SU(9)$ gauge field, the charged scalar fields we introduce to effect symmetry-breaking modify the 't Hooft vertex by contributing to the instanton density. In Section \ref{sec:interBreaking}, the breaking $SU(9) \rightarrow SU(3)^2/\mathbb{Z}_3$ was achieved by the vev of a three-index symmetric tensor $\Phi^{\lbrace ABC \rbrace}$, which decomposes under an $SU(2)$ subgroup to contribute $n(\frac{3}{2})=1, n(1)=7, n(\half)=28, n(0)=84$. In Section \ref{sec:flavorBreaking}, our toy model of flavor-breaking utilized $N_s=2$ scalar adjoints $\Sigma^A_{1,2}$ which contribute the same way as the vector, and a single scalar fundamental which contributes oppositely to one of the quarks.

Then we end up with a bottom yukawa interaction, after the extra fermion modes have been taken care of, which we can write in the form of an integral over the instanton scale with an infrared cutoff $\Lambda_9^{-1}$, 
\begin{equation} \label{eqn:inducedDownYuk}
    \mathcal{L} \supset y_t^\star e^{i\theta} \int_0^{\Lambda_9^{-1}} \frac{d\rho}{\rho} D(\rho) H \mathbf{Q} \mathbf{\bar d},
\end{equation}
and is sensitive both to the beta function at scales around $\Lambda_9$ and to the spectrum of charged matter.

Finally, to evaluate this scale integral, we must account for the non-trivial evolution of the gauge coupling as a function of the scale. We factor out the prefactor coming from the integration of quadratic fluctuations of charged fields, which does not depend on scale.
At one loop, the $\alpha$ evolution is $2\pi/\alpha(\rho) = 2\pi/\alpha(\Lambda_9) - \beta_9 \log (\rho \Lambda_9)$, leading to an integral over the size modulus which eventually looks like
\begin{equation}
    \int \frac{d\rho}{\rho} \left(\rho \Lambda_9\right)^{\beta_9} \left( \frac{2\pi}{\alpha(\Lambda_9)} - \beta_9 \log (\rho \Lambda_9)\right)^{2N}.
\end{equation}
The prefactor in the instanton density \eqref{eqn:instDens} is extremely small for large $N$, so it is unclear if large $y_b$ can be generated. But in fact the integral can be quite large, which is evident for large $\alpha$ as the integral approaches $(2N)!/\beta_9$, so can potentially compete with the prefactor. Then overall for large $N$, $(2N)!/(N-1)!(N-2)! \sim 2^{2N}\sqrt{N^5/\pi}$, and the small constant factors are indeed compensated for. 

\bibliographystyle{jhep}
\bibliography{strongcp}
\end{document}